\documentclass[10pt,aps,physrev,reprint,floatfix,superscriptaddress,longbibliography]{revtex4-2}
\pdfoutput=1
\usepackage{graphicx}
\usepackage[export]{adjustbox}
\usepackage{dcolumn}
\usepackage{bm}
\usepackage{graphicx}  
\usepackage{dcolumn}   
\usepackage{bm}        
\usepackage{verbatim}   
\usepackage[usenames]{color}  
\usepackage{stmaryrd}
\usepackage{float}
\usepackage{units}
\usepackage{textcomp}
\usepackage{textcomp}
\usepackage{amsmath}
\usepackage{commath}
\usepackage{amssymb}
\usepackage{esint}
\usepackage{ae,aecompl}
\usepackage[colorlinks=true,linkcolor=blue,citecolor=blue]{hyperref}
\usepackage{subfigure}
\begin{document}

\title{Magnetic properties of a cavity-embedded square lattice of quantum dots or antidots}

\author{Vram Mughnetsyan}
	\email{vram@ysu.am}
	\affiliation{Department of Solid State Physics, Yerevan State University, Alex Manoogian 1, 0025 Yerevan, Armenia}
\author{Vidar Gudmundsson}
	\email{vidar@hi.is}
	\affiliation{Science Institute, University of Iceland, Dunhaga 3, IS-107 Reykjavik, Iceland}
	\author{Nzar Rauf Abdullah}
	\affiliation{Physics Department, College of Science,
		University of Sulaimani, Kurdistan Region, Iraq}
	\affiliation{Computer Engineering Department, College of Engineering, Komar University
		of Science and Technology, Sulaimani 46001, Kurdistan Region, Iraq}
	\author{Chi-Shung Tang}
	\email{cstang@nuu.edu.tw}
	\affiliation{Department of Mechanical Engineering, National United University, Miaoli 36003, Taiwan}
	\author{Valeriu Moldoveanu}
	\email{valim@infim.ro}
	\affiliation{National Institute of Materials Physics, PO Box MG-7, Bucharest-Magurele,
		Romania}
	\author{Andrei Manolescu}
	\email{manoles@ru.is}
	\affiliation{Department of Engineering, Reykjavik University, Menntavegur
		1, IS-102 Reykjavik, Iceland}

\begin{abstract}
We apply quantum electrodynamical density functional theory to obtain the electronic density, the spin polarization, as well as the orbital and the spin magnetization of square periodic arrays of quantum dots or antidots subjected to the influence of a far-infrared cavity photon field. A gradient-based exchange-correlation functional adapted to a two-dimensional electron gas in a transverse homogeneous magnetic field is used in the theoretical framework and calculations. The obtained results predict a non-trivial effect of the cavity field on the electron distribution in the unit cell of the superlattice, as well as on the orbital and the spin magnetization. The number of electrons per unit cell of the superlattice is shown to play a crucial role in the modification of the magnetization via the electron-photon coupling. The calculations show that cavity photons strengthen the diamagnetic effect in the quantum dots structure, while they weaken the paramagnetic effect in an antidot structure. As the number of electrons per unit cell of the lattice increases the electron-photon interaction reduces the exchange forces that would otherwise promote strong spin splitting for both the dot
and the antidot array.
\end{abstract}
\maketitle

\section{INTRODUCTION}
The rich physics behind the properties of periodic structures in a homogeneous magnetic field have been attracting a huge interest of scientists since the appearance of the works of Azbel \cite{Azbel} and Hofstadter \cite{Hofstadter}. The study of electron gas properties in a two-dimensional (2D) periodic lattice subjected to a transverse (perpendicular to the lattice plane) and homogeneous magnetic field reopens new perspectives in applied science due to the technology achievements in the field of manufacturing of high quality ordered quantum dot structures in two dimensions. The rich structure of the energy spectrum in such structures in magnetic field is connected to the commensurability of the magnetic length and the lattice constant of the superlattice structure \cite{Gudmundsson0,Gumbs,Guil,Zhang0}. The fractal structure of the Landau band splitting have primarily been observed using the Peierls' substitution \cite{Hofstadter} in calculations of the energy dispersion for an electron in a superlattice (SL). An other method which has been used successfully is the direct diagonalization of the Hamiltonian with an appropriate choice of basis functions in which magnetic translational phases are included. Both the Landau \cite{Pfannkuche} and the symmetric \cite{Ferrari, Silberbauer} gauges for the vector potential are shown to provide results with high accuracy.

The study of optical properties of two-dimensional electron systems (2DES) \cite{Yoshie,Zhang,Stockklauser},
quasiparticle excitation in light-matter systems \cite{Savona,Dini,Ciuti,Kyriienko}, or processes
in chemistry \cite{Hutchison,Garcia-Vidal,Thomas,Schafer1}, has been gaining attention in the last three decades. The diverse systems and their phenomena have been theoretically described by a multitude of methods ranging from simple toy models \cite{Jaynes,Bishop}, nonequilibrium Green’s functions \cite{Torre}, and master equations of various types. In the more complex
models with few to many charged entities, traditional
approaches to many-body theory, or configuration interactions (exact numerical diagonalization in a truncated many-body Fock space), have been used \cite{Moldoveanu}, but relatively recently, density functional theory
approaches have been appearing \cite{Ruggenthaler,Schafer2,Flick00,Flick0}.
The magnetization of the low dimensional electron systems provides important information about the electronic ground state properties
and is one of the most fundamental properties of matter which is governed by many-body effects \cite{Meinel,Schwarz}. It is especially adequate to explore in future experiments as it is an equilibrium measurable quantity. In particular, the orbital magnetization may be a useful quantity to test the quality of the exchange and correlation functionals for the electron-photon
interactions.

In this paper we present a comparative study of the magnetic properties of square SLs composed of quantum dots and antidots embedded in a far-infrared photon cavity and subjected to a transverse and homogeneous magnetic field.
A quantum electrodynamical density functional theory is applied to obtain the electron density, the spin polarization, as well as the orbital and spin magnetization of the systems.
A gradient-based exchange-correlation functional adapted to the two-dimensional electron gas in a transverse
homogeneous magnetic field is used for theoretical framework and the calculations.

For three decades it has been known from experiments that in regular arrays of quantum dots
the same number of electrons tends to reside in each dot due to the Coulomb blockade \cite{Meurer92}.
In arrays of quantum antidots this ``charge quantization'' is not present and the full power of
electron screening that varies strongly with the fractional filling or occupation of energy bands
sets in. In order to compare the properties of arrays of dots and antidots we will thus allow for
a fractional mean electron number in each lattice unit, even though that is not a proper description
of dot arrays unless they are shallow or have a significant overlap of electron density between dots.

The paper is organized as follows: in section II the theoretical model is described, section III is devoted to the discussion on the cavity field effects on the electronic density and the spin polarization, in section IV the orbital and the spin magnetization as functions of the electrons' number and the magnetic field flux in the SL unit cell are discussed and finally in section V conclusions are presented.

\section{Theoretical model}
We consider an electron gas in a 2D square SL composed of QDs or AQDs. This structure can be modeled by a periodic modulating superlattice potential
\begin{equation}
\label{potential}
    V_{\mathrm{SL}}(\boldsymbol{r})=V_{0}\left[\sin \left(\frac{g_{1} x}{2}\right) \sin \left(\frac{g_{2} y}{2}\right)\right]^{2}
\end{equation}
with $V_{0}= \pm 16.0 \mathrm{meV}$. In Eq.\ (\ref{potential}) a negative value of $V_{0}$ corresponds to QD SL, while a positive value stands for an AQD SL. The primitive vectors of the reciprocal lattice, $\boldsymbol{g}_{1(2)}=g_{1(2)}\hat{\boldsymbol{e}}_{x(y)}$, where $g_{1(2)}=2\pi/L_{1(2)}$, and $L_{1(2)}$ are the direct lattice constants along the unit vectors of the Cartesian coordinates
$\hat{\boldsymbol{e}}_{x}$ and $\hat{\boldsymbol{e}}_{y}$, respectively.

The Coulomb interactions of the electrons is taken into account in the framework of local spin-density functional theory (LSDFT) in the presence of a transverse homogeneous magnetic field $\boldsymbol{B}=B \hat{\boldsymbol{e}}_{z}$ \cite{Gudmundsson1}.

In the framework of the QEDFT the total one-electron Hamiltonian of the electrons in the periodic potential  (\ref{potential}) positioned in a photon cavity can be expressed as follows
\begin{equation}
\label{totalH}
      H=H_{0}+H_{\mathrm{Z}}+V_{\mathrm{H}}+V_{\mathrm{xc}}+V_{\mathrm{xc}}^{\mathrm{EM}},
\end{equation}
where
\begin{equation}
\label{noninteractingH}
H_{0}=\frac{1}{2 m^{*}} \left(\boldsymbol{p}+\frac{e}{c} \boldsymbol{A}\right)^{2}+V_{\mathrm{SL}}(\boldsymbol{r})
\end{equation}
is the one-electron Hamiltonian in the SL in the magnetic field, $H_{\mathrm {Z}}= \pm g^{*} \mu_{\mathrm{B}}^{*} B / 2$ is the Zeeman term with the effective Land\'{e} factor $g^{*}$, and the Bohr magneton $\mu_{\mathrm{B}}^{*}$.
The Hartree-type Coulomb interaction is
\begin{equation}
V_{\mathrm{H}}(\boldsymbol{r})=\frac{e^{2}}{\kappa} \int_{\boldsymbol{R}^{2}} d \boldsymbol{r}^{\prime} \frac{\Delta n\left(\boldsymbol{r}^{\prime}\right)}{\left|\boldsymbol{r}-\boldsymbol{r}^{\prime}\right|},
\end{equation}
where $\Delta n(\boldsymbol{r})=n(\boldsymbol{r})-n_{\mathrm{b}}$, is the deviation of the electrons' density $n(\boldsymbol{r})$ from the homogeneous density of the background positive charge $n_{\mathrm{b}}$. The positive background charge guarantees the neutrality of the total system.
The last two terms of (\ref{totalH}) describe the exchange and correlation potentials connected with the Coulomb interaction between electrons $V_{\mathrm{xc}}$, and with the coupling to the cavity photons $V_{\mathrm{xc}}^{\mathrm{EM}}$.

For the iterative solution procedure we start from the Kohn-Sham eigenfunctions of the Hamiltonian (\ref{noninteractingH}), which had been constructed by Ferrari \cite{Ferrari} in the symmetric gauge for the vector potential, $\boldsymbol{A}=$ $(B / 2)(-y, x)$, and has been applied previously for band structure calculations by Silberbauer \cite{Silberbauer}. This choice of wavefunctions for the non-interacting electrons fulfills the commensurability conditions of the SL period and the magnetic length $l=(\hbar c /(e B))^{1 / 2}$ and reflects the periodicity of the system in a homogeneous magnetic field.

We denote the wave functions of the Ferrari basis \cite{Ferrari} by $\phi_{n_{l}}^{\mu \nu}(\boldsymbol{r})$ where $n_{l}=0,1,2, \cdots$ is the number of a Landau band, $\mu=\left(\theta_{1}+2\pi n_{1}\right) / p$ and $\nu=\left(\theta_{2}+2 \pi n_{2}\right) / q$, with $n_{1} \in I_{1}=\{0, \ldots, p-1\}$, $n_{2} \in I_{2}=\{0, \ldots, q-1\}$, and $\theta_{i} \in[-\pi, \pi]$. Here, $pq$ is the number of magnetic flux quanta $\Phi_{0}=h c / e$ flowing through a unit cell of the lattice.
Let us denote the eigenfunctions of the total Hamiltonian (\ref{totalH}) with $\psi_{\beta\bm{\theta}\sigma}(\boldsymbol{r})$, where $\sigma$ stands for the $z$ component of the spin, $\boldsymbol{\theta}=\left(\theta_{1}, \theta_{2}\right)$ defines a point in the magnetic Brillouin zone, and $\beta$ stands for all remaining quantum numbers. In each point of the reciprocal space $\boldsymbol{\theta}$ satisfying the condition $(\mu, \nu) \neq(\pi, \pi)$, the eigenfunctions $\phi_{n_{l}}^{\mu \nu}$ and $\psi_{\beta\bm{\theta}\sigma}$ form complete orthonormal sets.

The electron density can be presented as a sum of the corresponding densities with spin-up and spin-down states as follows
\begin{equation}
\label{density}
    \begin{aligned}
          n_{e}(\boldsymbol{r}) & =n_{\uparrow}(\boldsymbol{r})+n_{\downarrow}(\boldsymbol{r}) \\
          & =\frac{1}{(2 \pi)^{2}} \sum_{\boldsymbol{\beta},\sigma} \int_{-\pi}^{\pi} d \boldsymbol{\theta}\left|\psi_{\beta\bm{\theta}\sigma}(\boldsymbol{r})\right|^{2} f\left(E_{\beta{\bm\theta}\sigma}-\mu\right),
    \end{aligned}
\end{equation}
where the $\boldsymbol{\theta}$-integration is over the magnetic Brillouin zone, $f$ is the Fermi-Dirac distribution function with the chemical potential $\mu$, and $E_{\beta\bm{\theta}\sigma}$ is the energy spectrum of the Hamiltonian (\ref{totalH}).

The potential describing the exchange and correlation for the Coulomb interaction between the electrons
\begin{equation}
\label{xc}
   V_{\mathrm{xc}, \sigma}(\boldsymbol{r}, B)=\left.\frac{\partial}{\partial n_{\sigma}}\left(n_{e} \epsilon_{\mathrm{xc}}\left[n_{\uparrow}, n_{\downarrow}, B\right]\right)\right|_{n_{\sigma}=n_{\sigma}(r)},
\end{equation}
has been previously derived from the Coulomb exchange and correlation functionals in \cite{Gudmundsson1}. In \cite{Gudmundsson1} it is also shown how the exchange and correlation functional suggested by Johannes Flick \cite{Flick} can be adapted for a 2DEG in a perpendicular magnetic field resulting in the following expression
\begin{align}
\label{xcEM_E}
E_{\mathrm{xc}}^{\mathrm{GA}}[n,\bm{\nabla}n]= & \frac{1}{16\pi} \sum_{\alpha=1}^{N_{p}}\left|\lambda_{\alpha}\right|^{2} \\
& \times \int d \boldsymbol{r} \frac{\hbar \omega_{p}(\boldsymbol{r})}{\sqrt{\left(\hbar \omega_{p}(\boldsymbol{r})\right)^{2} / 3+\left(\hbar \omega_{g}(\boldsymbol{r})\right)^{2}}+\hbar \omega_{\alpha}},\nonumber
\end{align}
where $\hbar \omega_{\alpha}$ and $\lambda_{\alpha}$ are the energy and the coupling strength of cavity-photon mode $\alpha$, respectively, and $N_{p}$ is the number of cavity modes. The gap-energy  \cite{Vydrov1,Flick,Vydrov2} stemming from considerations of dynamic dipole polarizability leading to the van der Waals interaction is given by
\begin{equation}
\label{gap}
   \left(\hbar \omega_{g}(\boldsymbol{r})\right)^{2}=C\left|\frac{\bm{\nabla} n_{e}}{n_{e}}\right|^{4} \frac{\hbar^{2}}{m^{* 2}}=C\left(\hbar \omega_{c}\right)^{2} l^{4}\left|\frac{\bm{\nabla} n_{e}}{n_{e}}\right|^{4},
\end{equation}
with $C=0.0089$ and the cyclotron frequency $\omega_{c}=$ $e B /\left(m^{*} c\right)$. The dispersion of a 2D magnetoplasmon is \cite{Stern,Ando}
\begin{equation}
\label{magnetoplasmon_q}
 \left(\hbar \omega_{p}(q)\right)^{2}=\left(\hbar \omega_{c}\right)^{2}+\frac{2 \pi n_{e}^{2}}{m^{*} \kappa} q+\frac{3}{4} v_{\mathrm{F}}^{2} q^{2},
 \end{equation}
where $v_{\mathrm{F}}$ is the Fermi velocity, and $q$ is a general wave vector. Importantly, the gap of the plasmon is only due to magnetic field, which indicates that the $2 \mathrm{DEG}$ is softer, regarding external perturbation, than a 3D electron gas. Using the relation $q \approx k_{\mathrm{F}} / 6 \approx\left|\bm{\nabla} n_{e}\right| / (6n_{e})$ \cite{Perdew,Langreth} we obtain a local plasmon dispersion as follows
\begin{equation}
\label{magnetoplasmon_r}
\begin{aligned}
\left(\hbar \omega_{p}(\boldsymbol{r})\right)^{2}= & \left(\hbar \omega_{c}\right)\left(2 \pi l^{2} n_{e}(\boldsymbol{r})\right)\left(\frac{e^{2}}{\kappa l}\right)\left(l \frac{\left|\bm{\nabla} n_{e}\right|}{6n_{e}}\right) \\
& +\left(\hbar \omega_{c}\right)^{2}\left\{\frac{1}{36}\left(\frac{\left|l \bm{\nabla} n_{e}\right|}{n_{e}}\right)^{4}+1\right\}.
\end{aligned}
\end{equation}
We repeat here (\ref{magnetoplasmon_r}) as a small misprint has entered the earlier expression in
Ref.\ \onlinecite{Gudmundsson1} regarding the approximation $q\approx k_{\mathrm{F}} /6$.
Thus, it is clear that for the 2DES we need all the terms listed for the magnetoplasmon in order to have the treatment of the gap energy and the magnetoplasmon to the same order in $\bm{\nabla}n_e/n_e$. The exchange and the correlation potentials for the electron-photon interaction are then obtained by the variation
\begin{equation}
\label{xcEM_V}
V_{\mathrm{xc}}^{\mathrm{EM}}=\frac{\delta E_{\mathrm{xc}}^{\mathrm{GA}}}{\delta n_{e}}=\left\{\frac{\partial}{\partial n_{e}}-\bm{\nabla} \cdot \frac{\partial}{\partial\bm{\nabla} n_{e}}\right\} E_{\mathrm{xc}}^{\mathrm{GA}},
\end{equation}
together with the common extension to spin densities \cite{Perdew}
\begin{equation}
\label{Exc_spindens}
 E_{\mathrm{xc}}^{\mathrm{EM}}\left[n_{\uparrow}, n_{\downarrow}\right]=\frac{1}{2} E_{\mathrm{xc}}^{\mathrm{EM}}\left[2 n_{\uparrow}\right]+\frac{1}{2} E_{\mathrm{xc}}^{\mathrm{EM}}\left[2 n_{\downarrow}\right]
\end{equation}
and
\begin{equation}
\label{delExcdelnsigma}
\frac{\delta E_{\mathrm{xc}}^{\mathrm{EM}}\left[n_{\uparrow}, n_{\downarrow}\right]}{\delta n_{\sigma}}=\left.\frac{\delta E_{\mathrm{xc}}^{\mathrm{EM}}\left[n_{e}\right]}{\delta n_{e}(\boldsymbol{r})}\right|_{n_{e}(\boldsymbol{r})=2 n_{\sigma}(\boldsymbol{r})}.
\end{equation}

The electron-photon exchange-correlation potentials (\ref{xcEM_V})
depend on the electron density $n_e$ and its gradient $\bm{\nabla}n_e$ in a complicated way
and are added to the DFT self-consistency iterations. This dependence, and the
bandstructure, of the system make the exchange and correlation functionals and their potentials
depend on the number of electrons, $N_e$ in each unit cell in a nontrivial way.
We note that the self-consistency requirement in our calculations brings into play not only the first- and second-, but also higher order effects of repeated single photon processes \cite{Flick}.

\section{Electronic density and spin polarization}
%
\begin{figure}
      \includegraphics[width=0.23\textwidth,bb= 70 50 375 275]{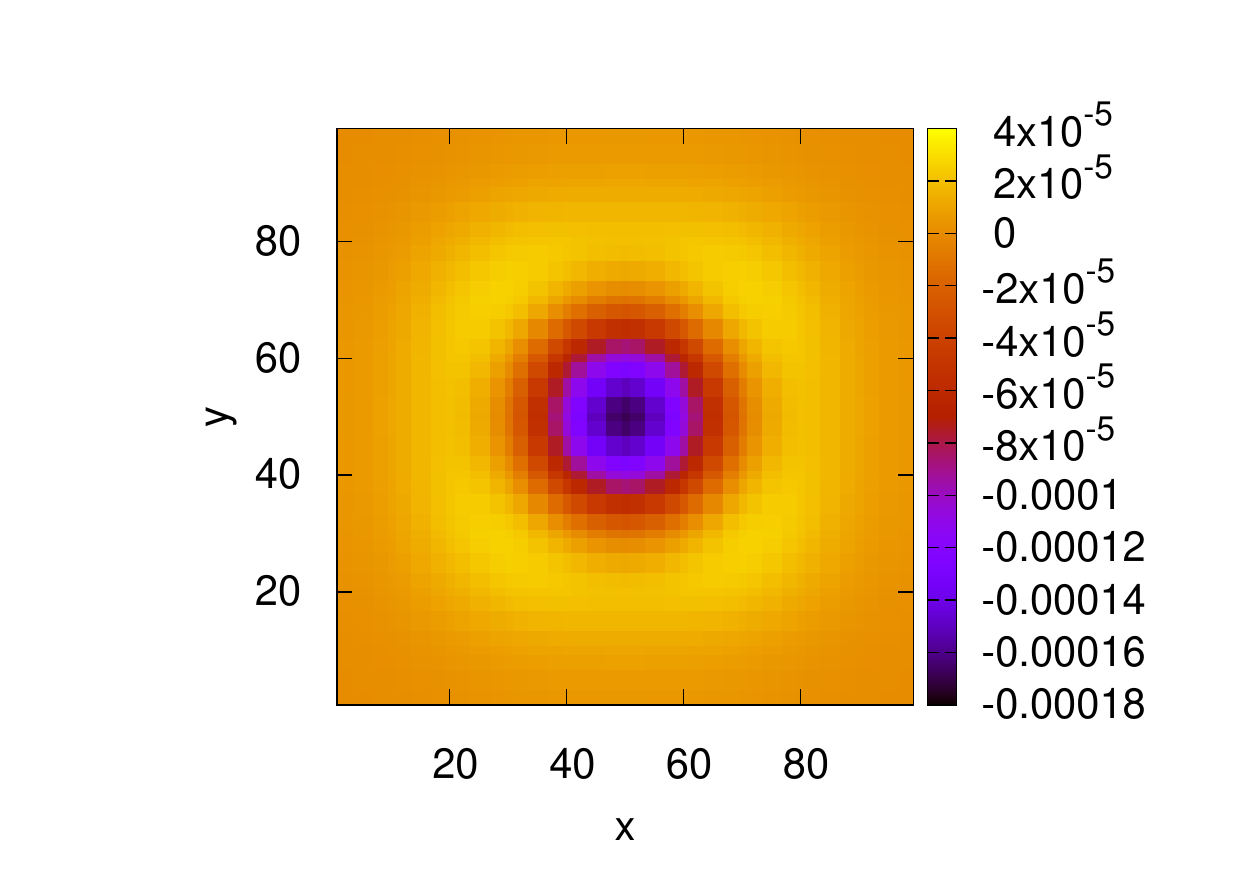}
      \includegraphics[width=0.23\textwidth,bb= 70 50 375 275]{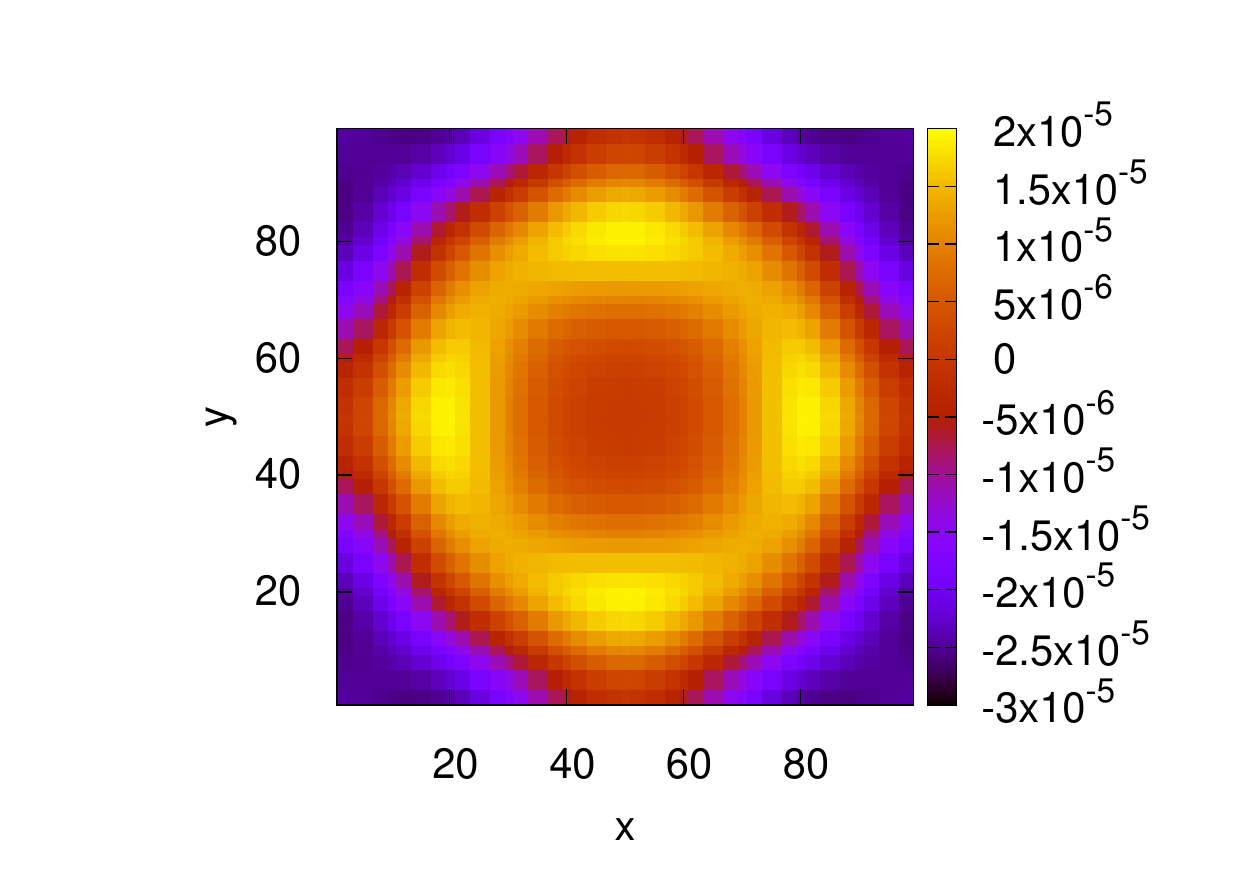}
      \includegraphics[width=0.23\textwidth,bb= 70 50 375 275]{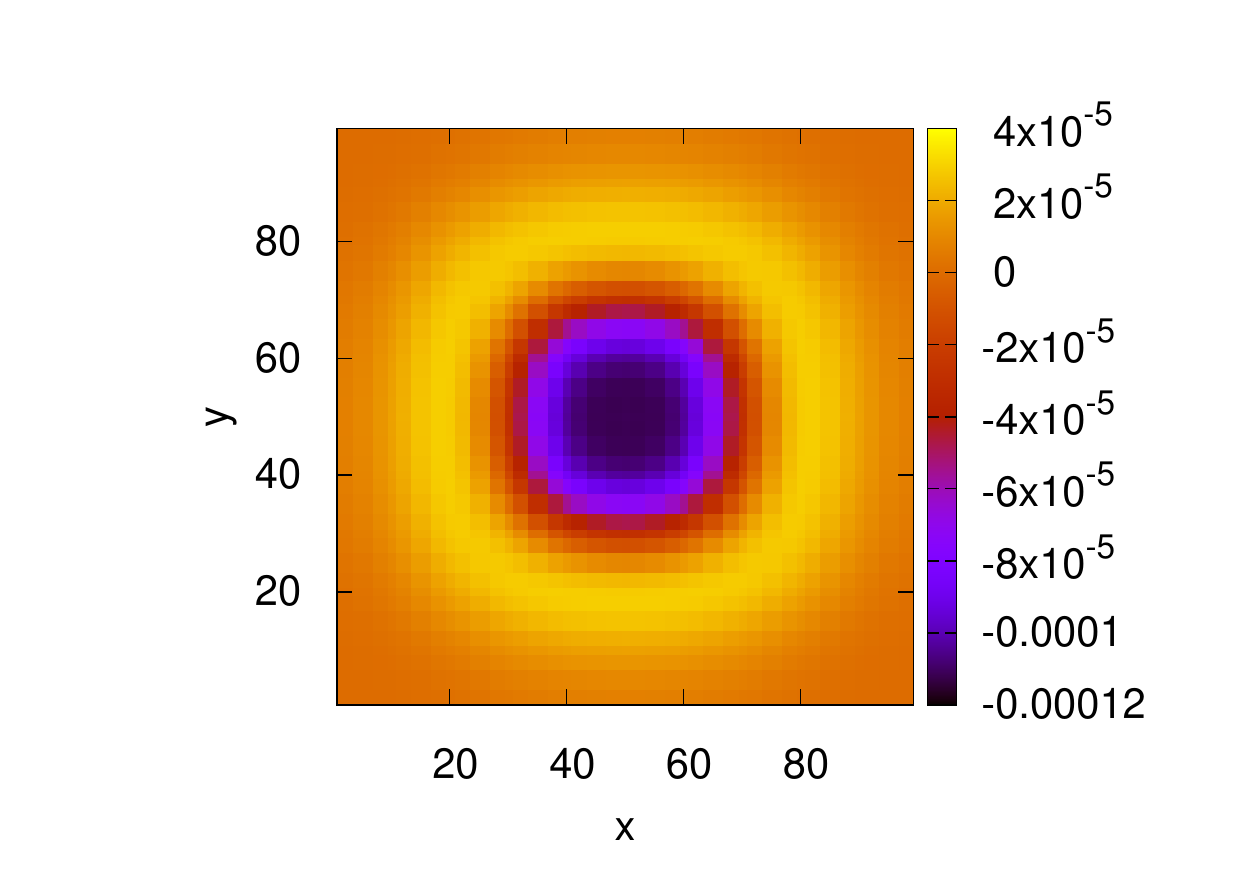}
      \includegraphics[width=0.23\textwidth,bb= 70 50 375 275]{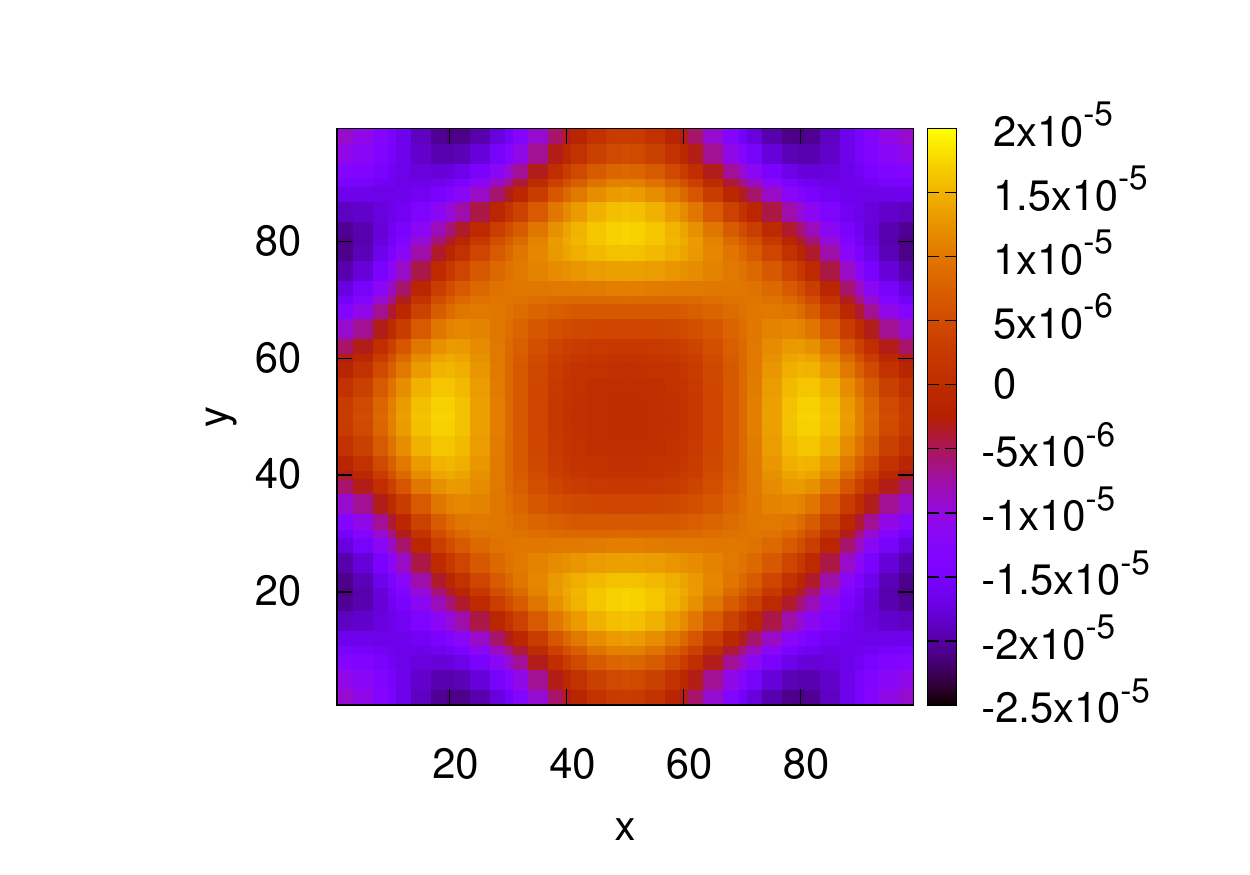}
      \includegraphics[width=0.23\textwidth,bb= 70 30 375 275]{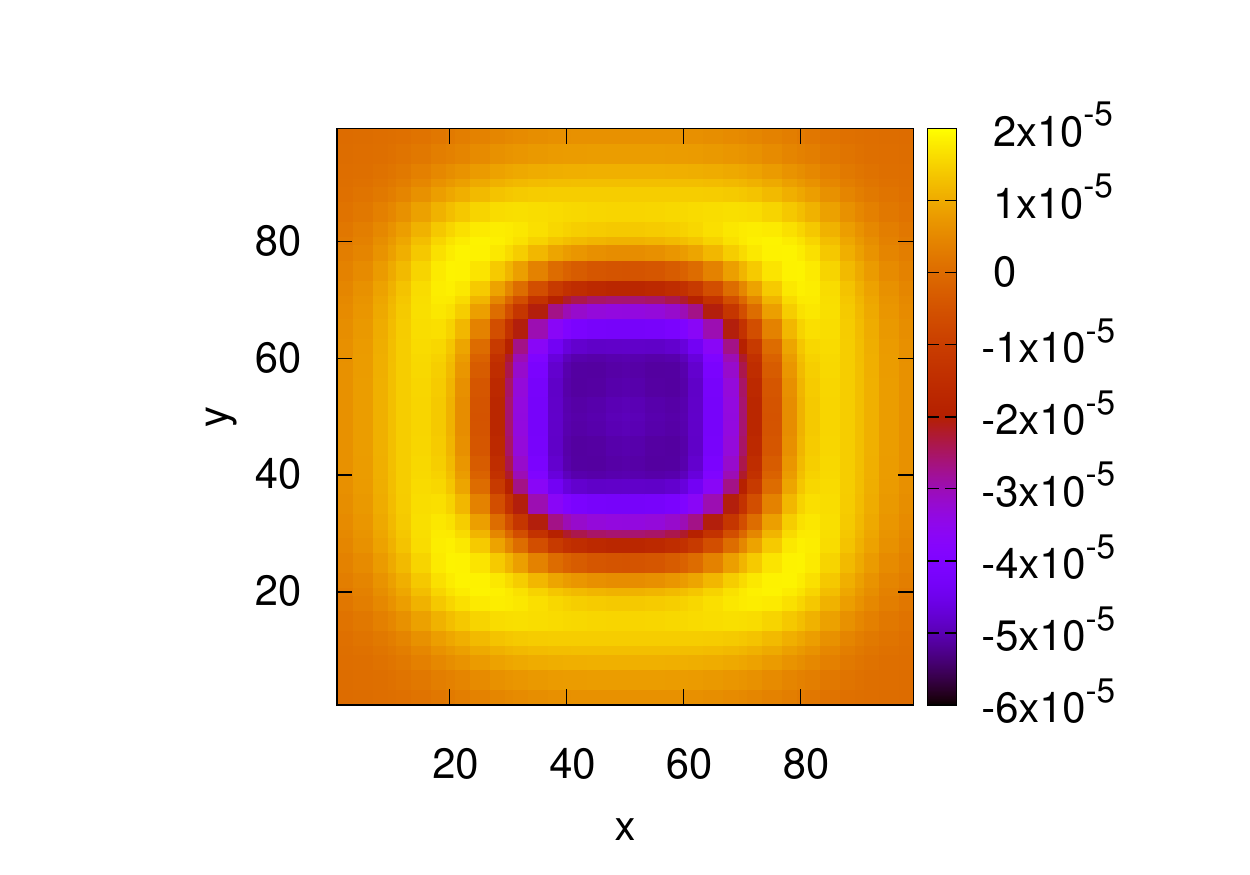}
      \includegraphics[width=0.23\textwidth,bb= 70 30 375 275]{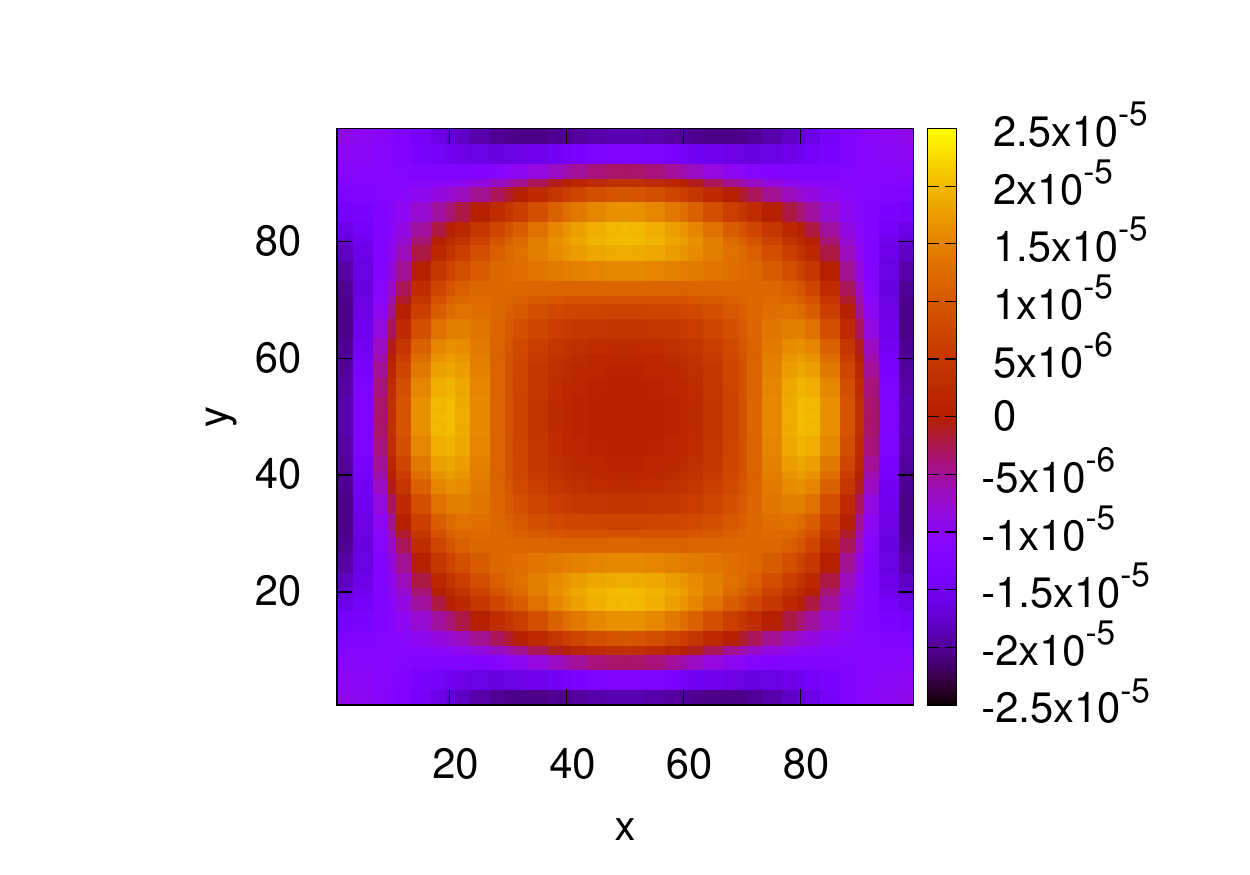}
      \caption{The change in the electron density due to the electron-photon coupling (EPC),
      $n_e(\lambda)-n_e(0)$ (in units of nm$^{-2}$),
      for $pq=2$, $N_e=2$ (1st row), $pq=4$, $N_e=2$ (2nd row), and $pq=3$, $N_e=2.2$ (3rd row). The left column corresponds to a QD SL, while the right one is for an AQD SL. $N_p=1$, $\lambda l= 0.1$ meV$^{1/2}$. }
      \label{Coupling_on_Density}
\end{figure}

\begin{figure}
      \includegraphics[width=0.23\textwidth,bb= 70 50 375 275]{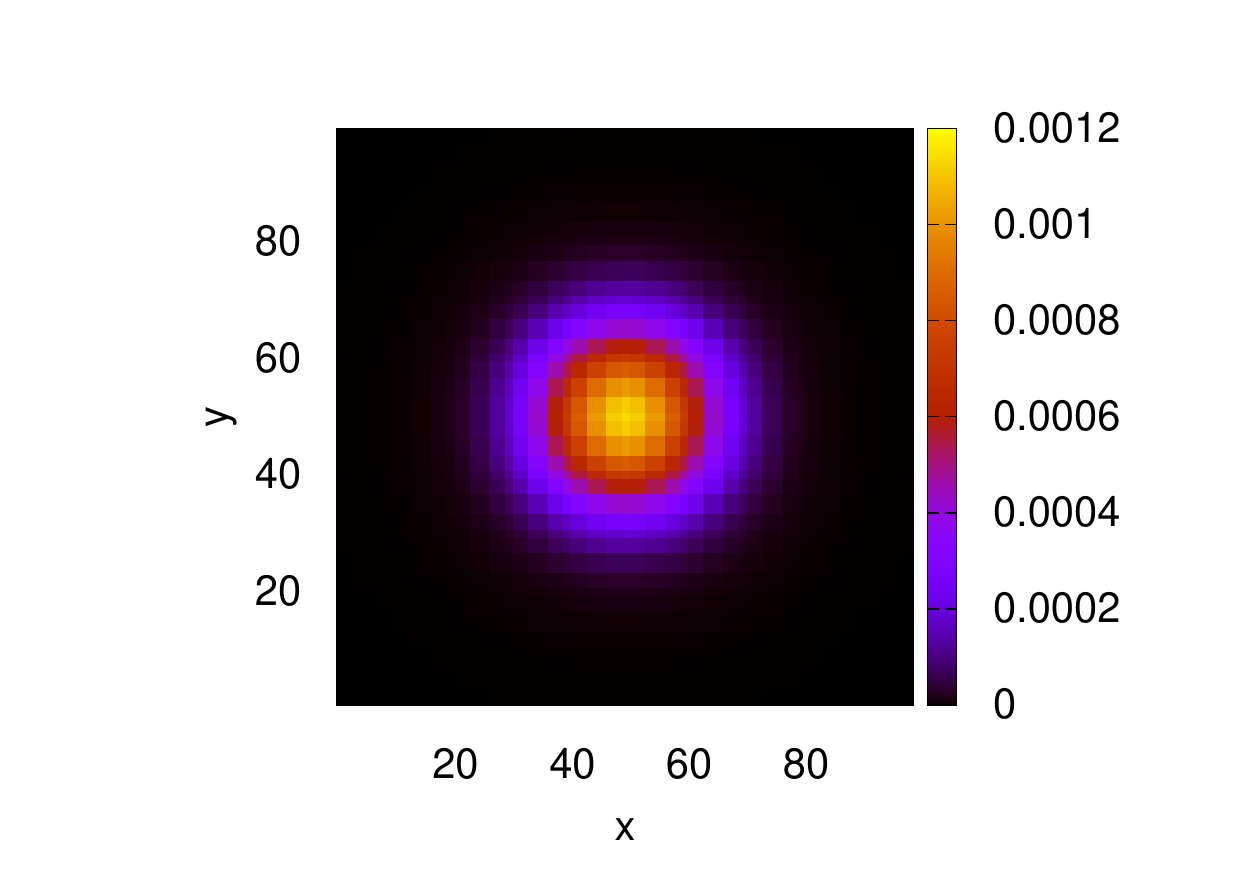}
      \includegraphics[width=0.23\textwidth,bb= 70 50 375 275]{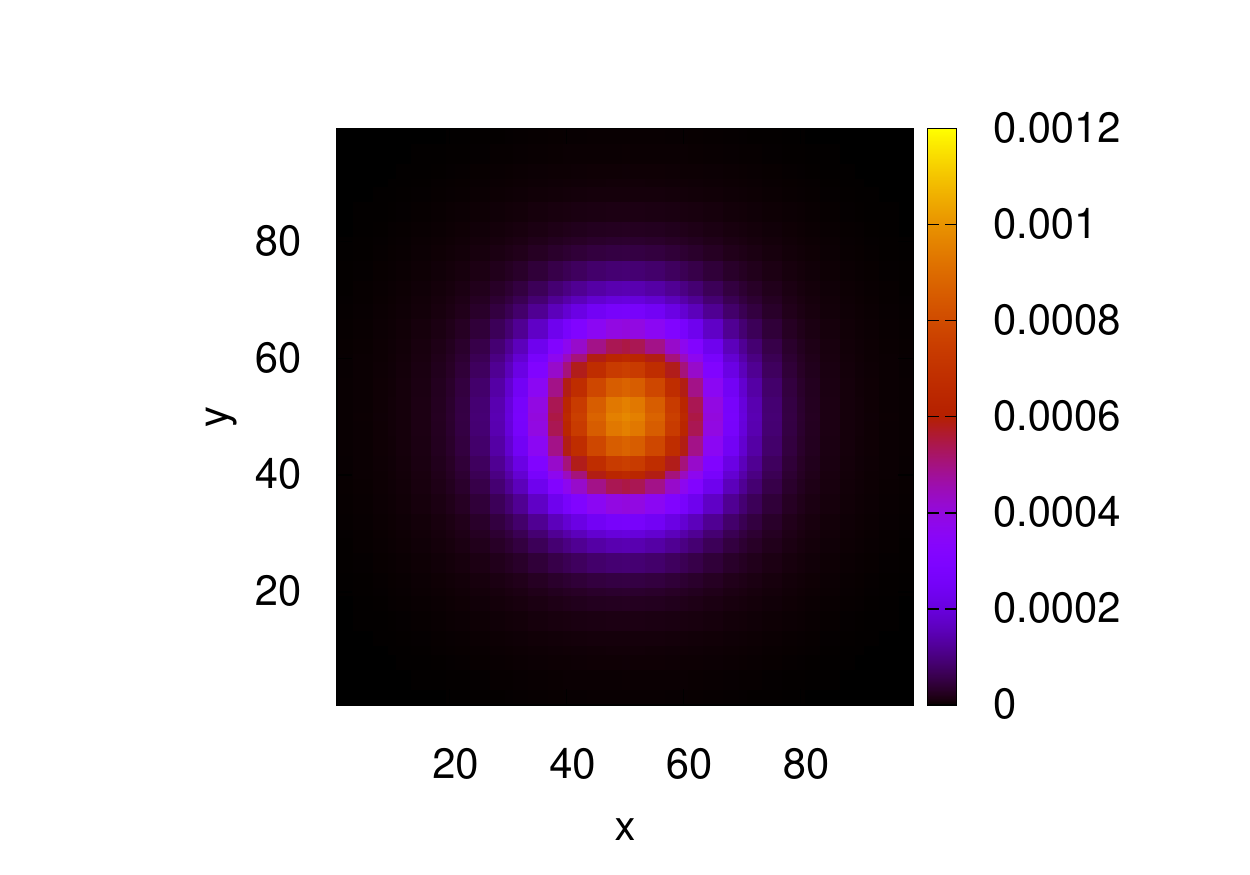}
      \includegraphics[width=0.23\textwidth,bb= 70 50 375 275]{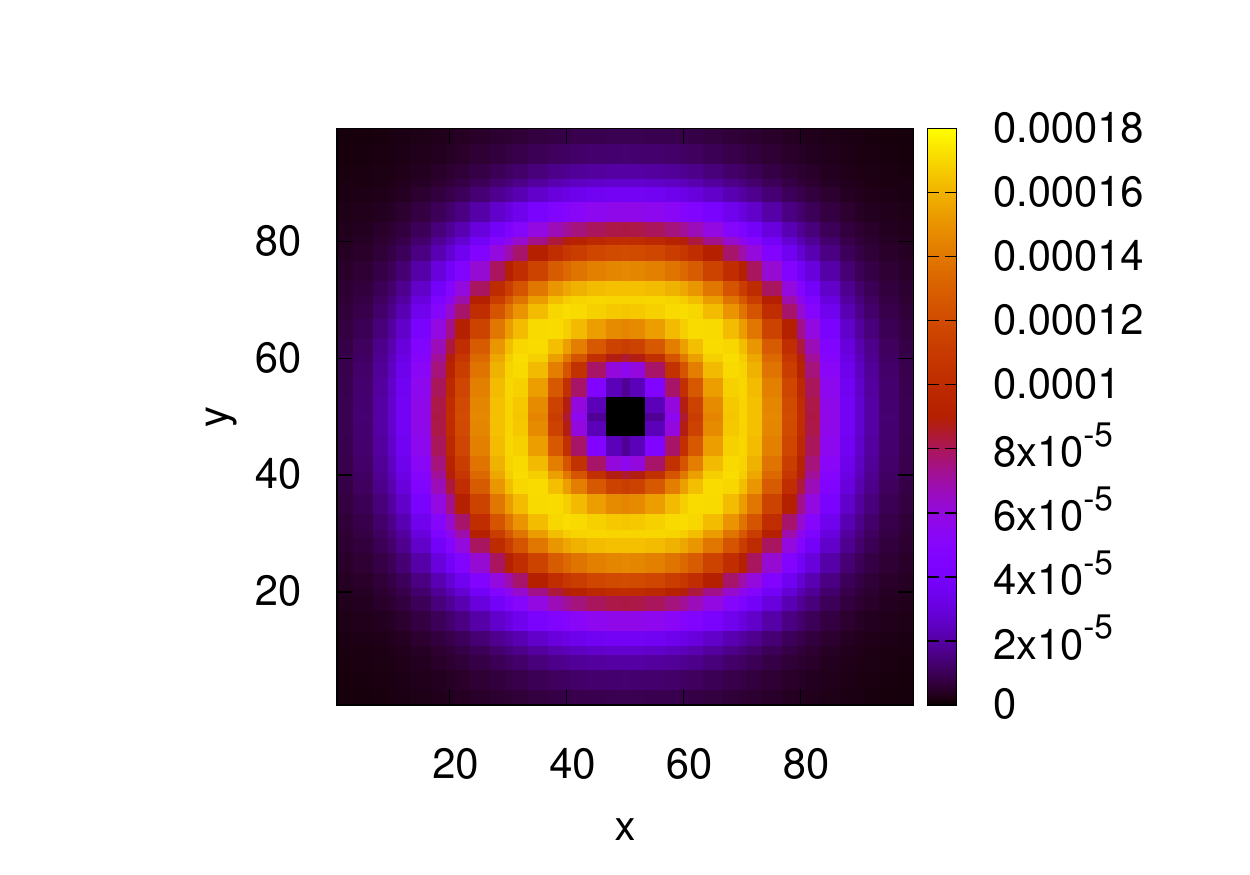}
      \includegraphics[width=0.23\textwidth,bb= 70 50 375 275]{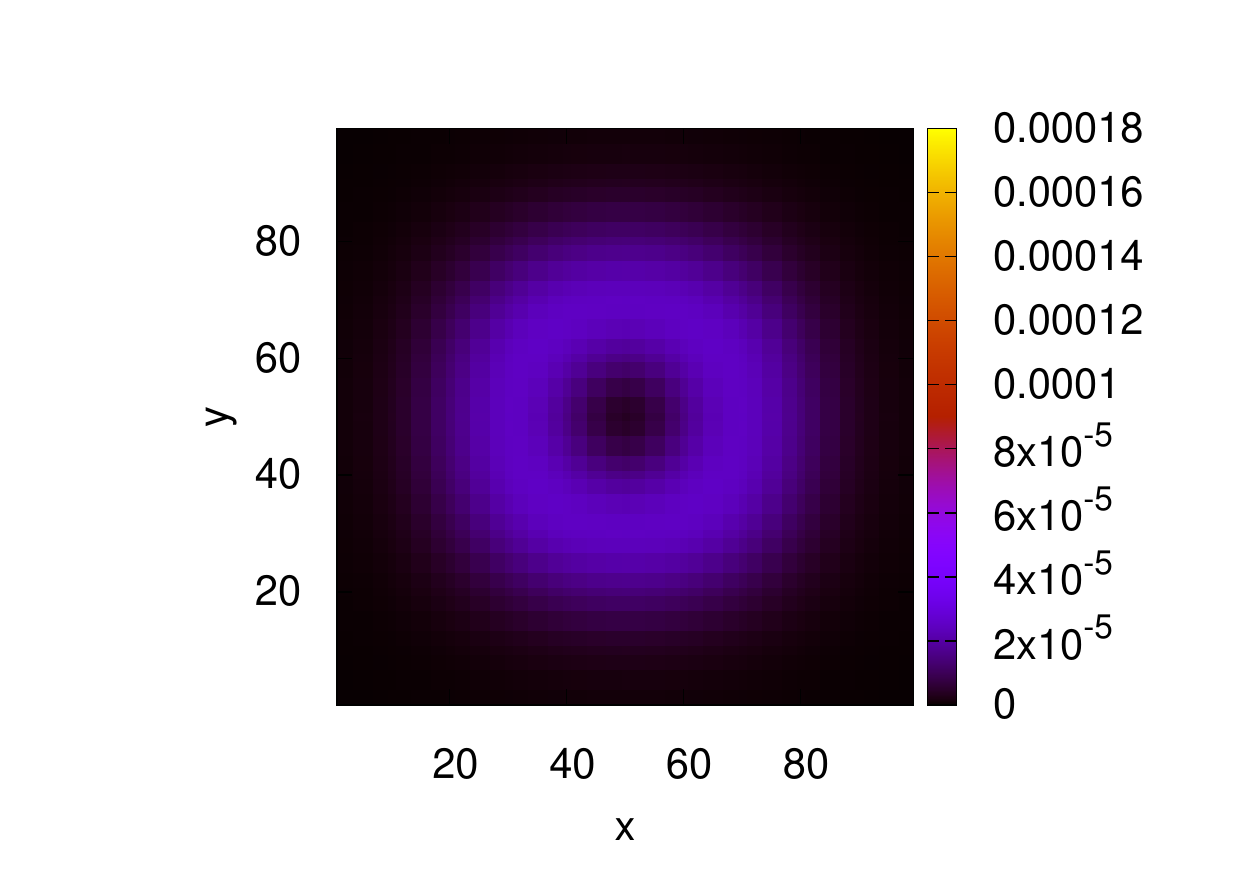}
      \includegraphics[width=0.23\textwidth,bb= 70 30 375 275]{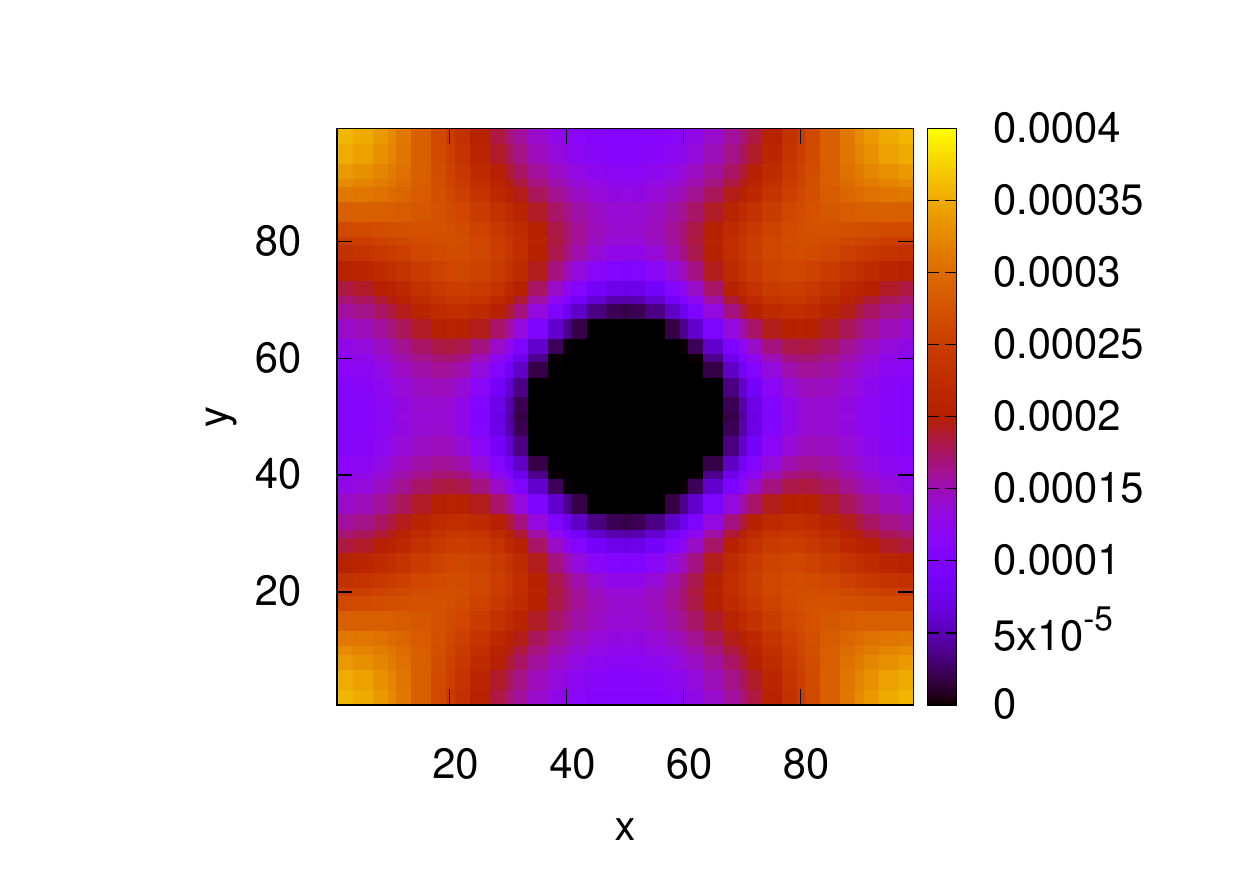}
      \includegraphics[width=0.23\textwidth,bb= 70 30 375 275]{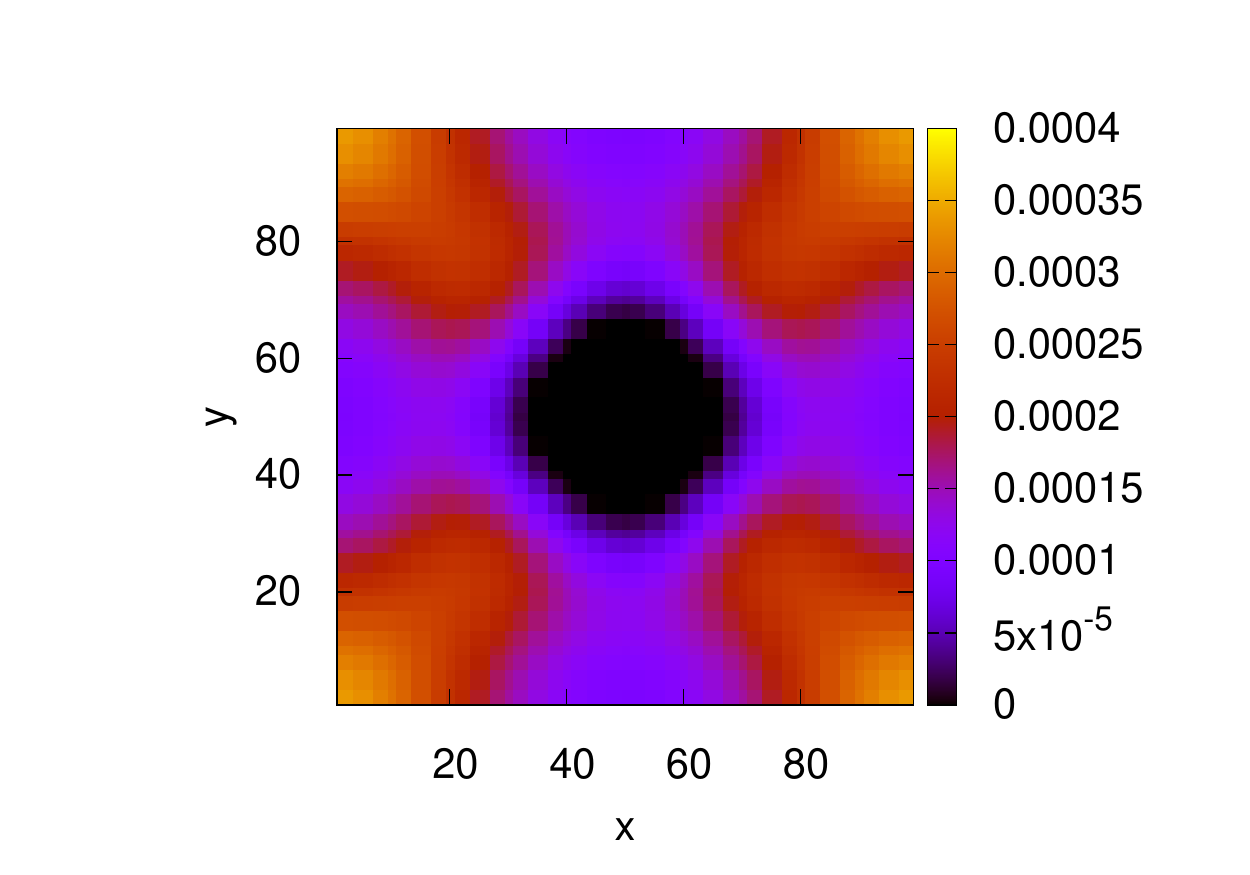}
      \caption{Spin polarization, $n_\uparrow - n_\downarrow$, (in units of nm$^{-2}$) in the absence of (left column), and the presence of (right column) the electron-photon coupling in a QD SL. The chosen values of parameters from the upper row to the lower one are $N_e=0.8$, $pq=3$, $N_e=2.6$, $pq=2$, and $N_e=6$, $pq=4$, respectively. $N_p=1$, $\lambda = 0.1$.}
     \label{Polarization_QD}
\end{figure}

\begin{figure}

         \includegraphics[width=0.23\textwidth,bb= 70 50 375 275]{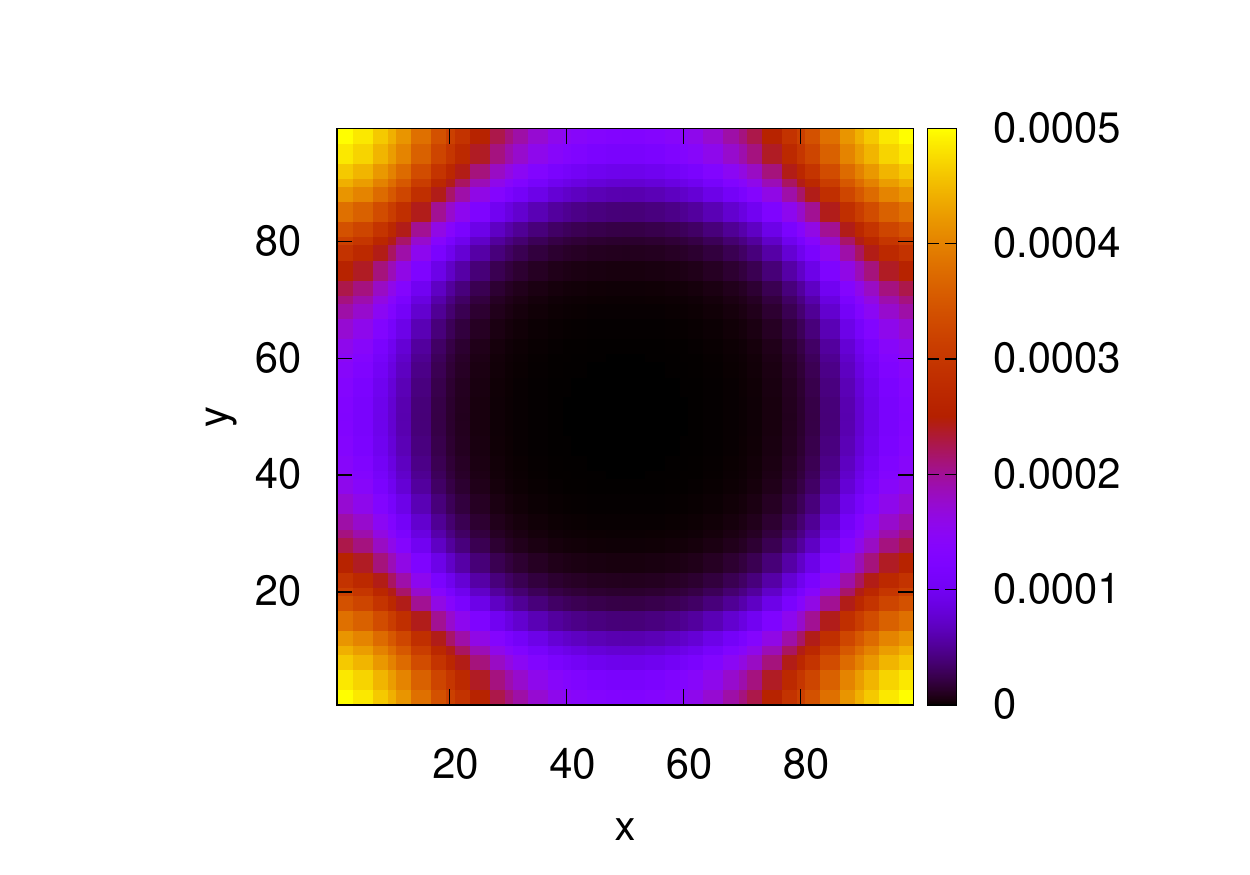}
         \includegraphics[width=0.23\textwidth,bb= 70 50 375 275]{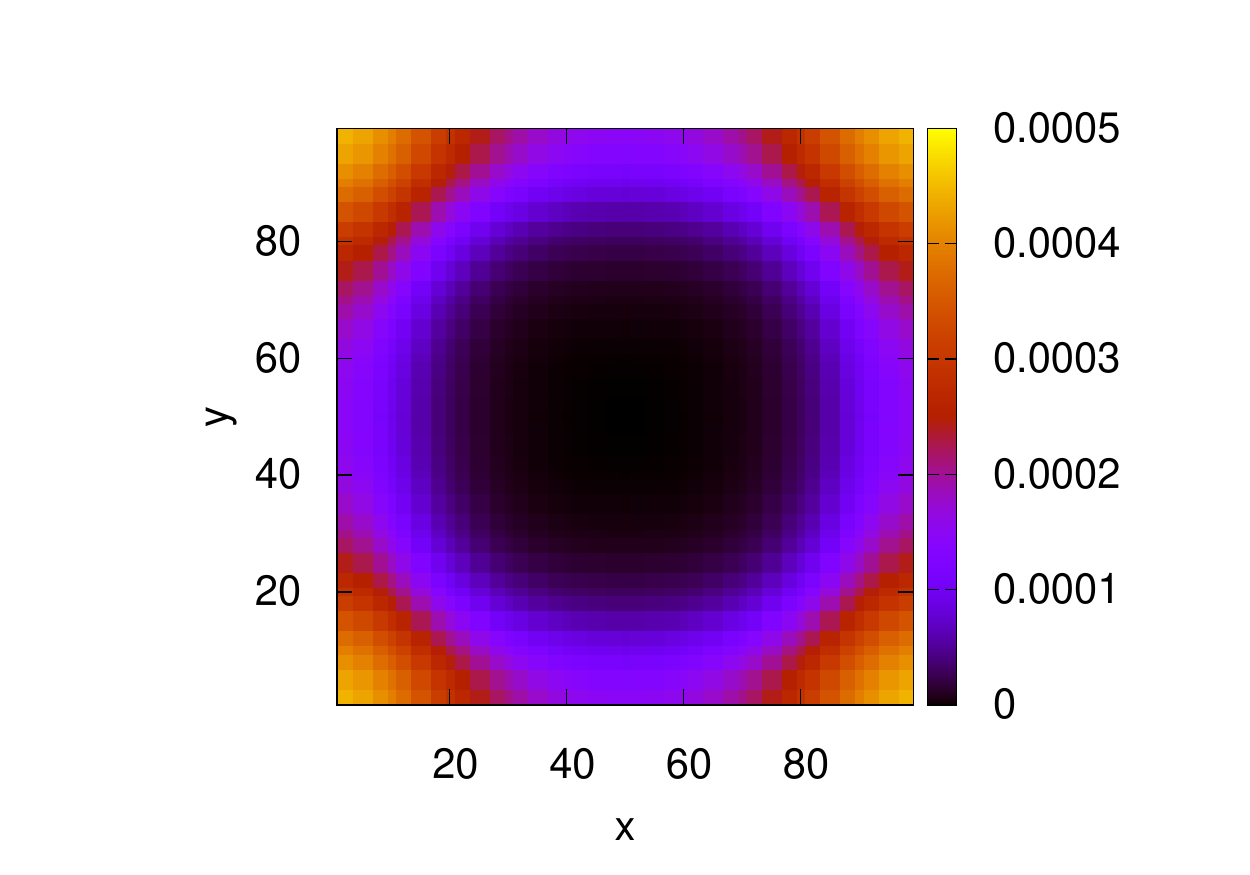}
         \includegraphics[width=0.23\textwidth,bb= 70 50 375 275]{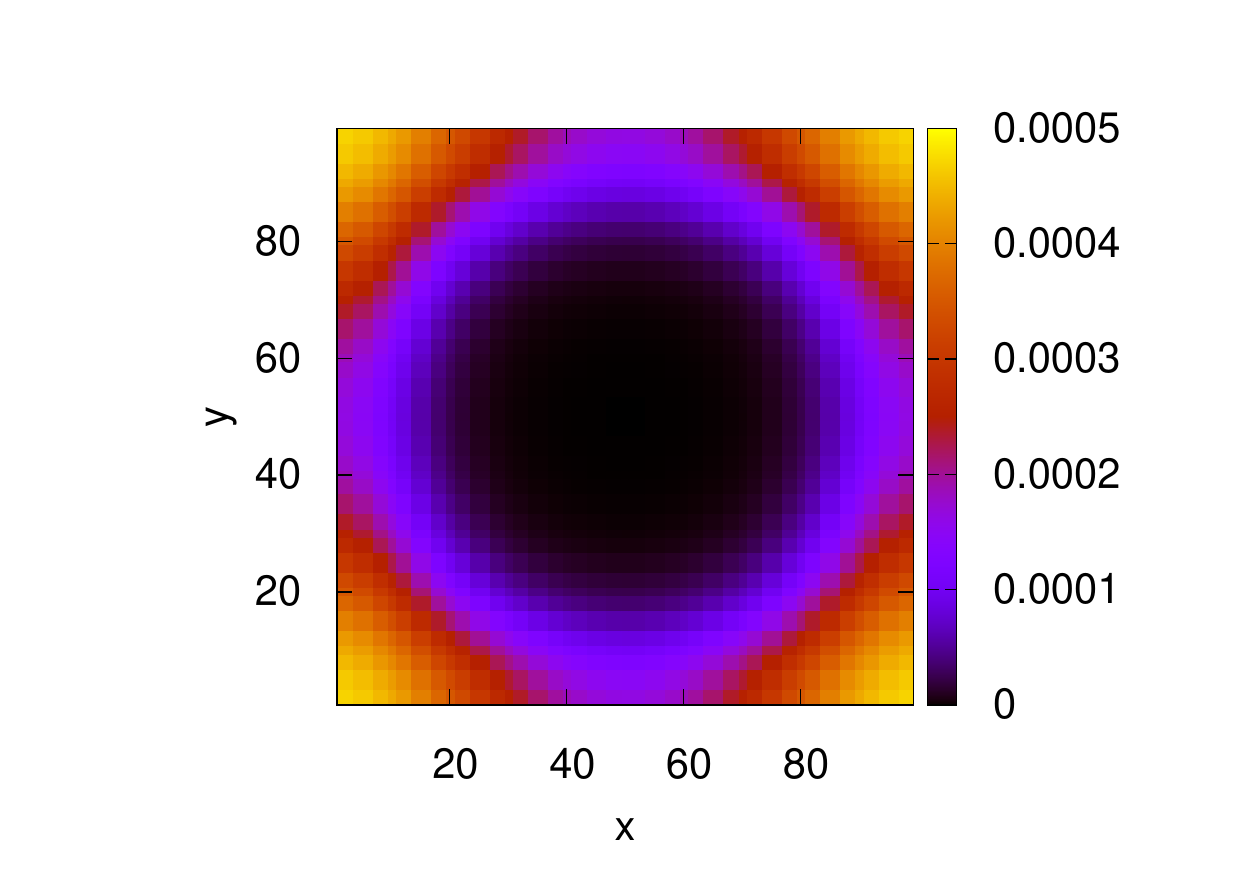}
         \includegraphics[width=0.23\textwidth,bb= 70 50 375 275]{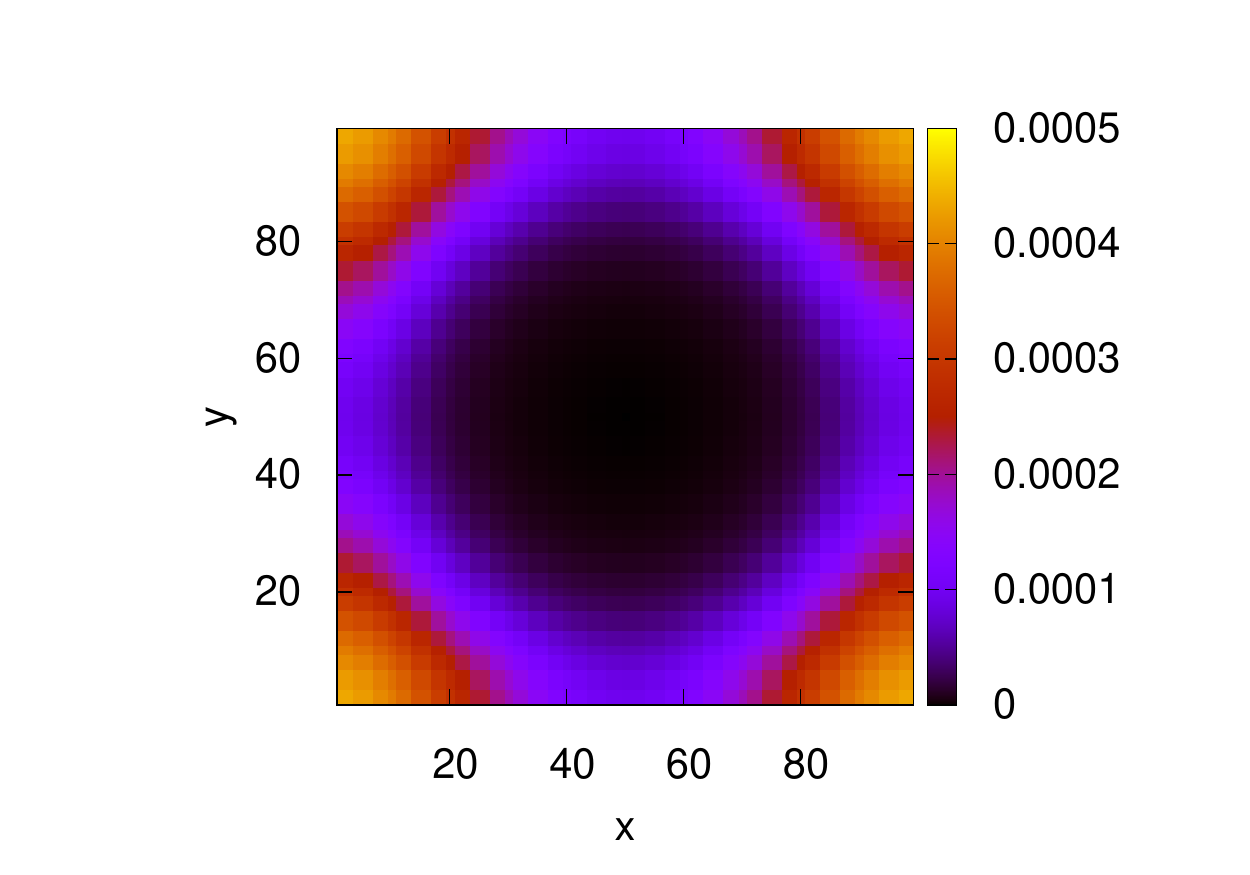}
         \includegraphics[width=0.23\textwidth,bb= 70 30 375 275]{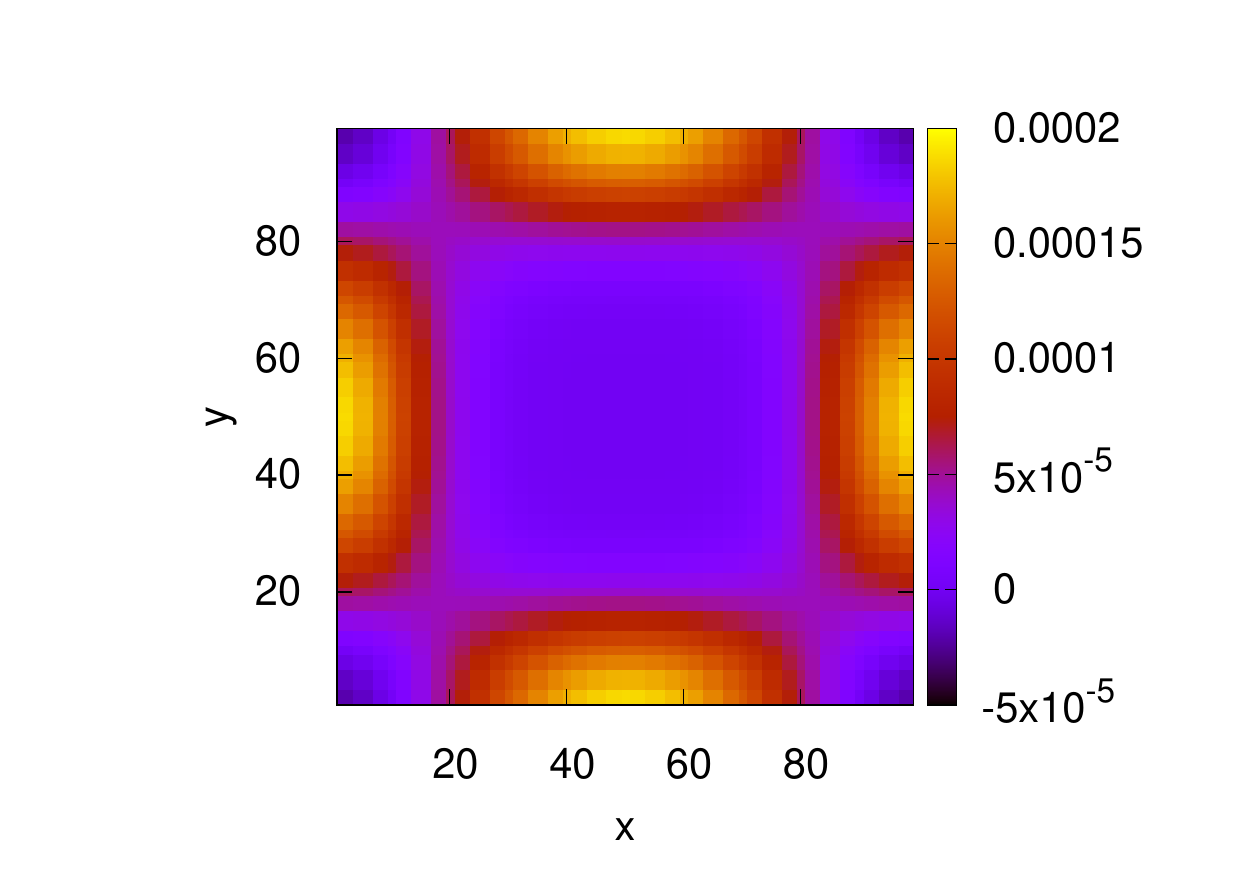}
         \includegraphics[width=0.23\textwidth,bb= 70 30 375 275]{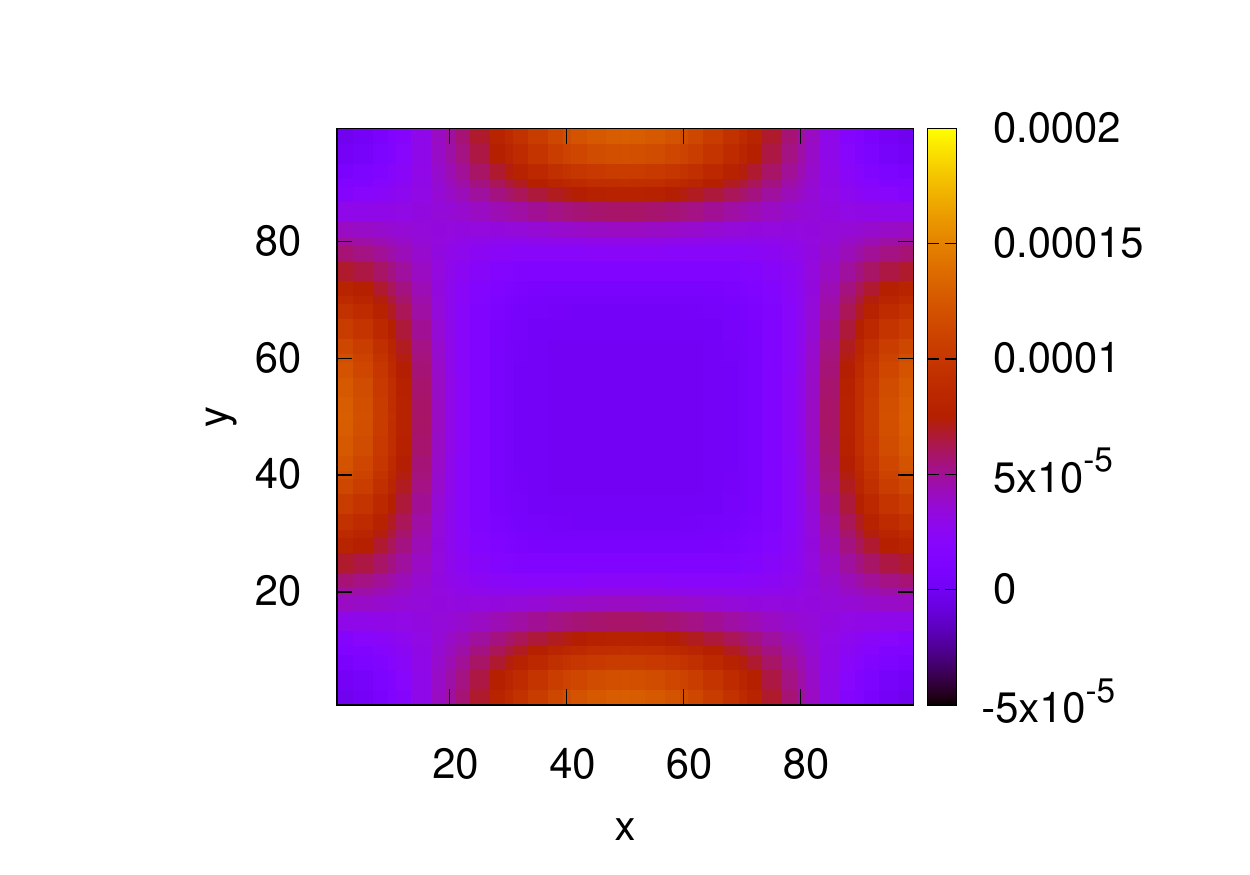}
        \caption{Spin polarization, $n_\uparrow - n_\downarrow$, (in units of nm$^{-2}$) in the absence of (left column), and the presence of (right column) the electron-photon coupling in the AQD SL. The chosen values of parameters from the upper row to the lower one are $N_e=1.2$, $pq=4$, $N_e=1.4$, $pq=3$, and $N_e=2.6$, $pq=2$, respectively.
       $N_p=1$, $\lambda = 0.1$.}
        \label{Polarization_AQD}
\end{figure}

In our calculations we choose the temperature $T=1.0 \mathrm{~K}$ and use GaAs parameters $m^{*}=0.067 m_{e}, \kappa=12.4$, and $g^{*}=0.44$. The lattice constants in both directions are the same: $L_{1}=L_{2}=L=100$nm.

The effects of the electron-photon coupling (EPC) on the electronic density, namely the difference of densities with and without the EPC, $n_{e}(\lambda)-n_e(0)$, for different values of magnetic flux and the number of electrons per UC of QD SL (the left column) and AQD SL (the right column) are presented in
Fig.\ \ref{Coupling_on_Density}
for coupling strength $\lambda l=0.1$ meV$^{1/2}$. It is obvious that the cavity field smears electronic density over the UC. As a result the density increases (decreases) in the regions of potential barriers (wells). Note, that for QD SL the wells are around the center of UC, while for AQD SL the center of each UC corresponds to potential barrier.
Another peculiarity of the EPC is that it is most significant in the regions of some intermediate distance from the center of the UC, both for the QD and the ADQ SLs, where the gradient terms of $n_el^2$ are large.
For the case of two electrons in a UC the increase in the magnetic flux leads to more distant regions of the maximum influence of the EPC in the QD SLs (compare the 1st and the 2nd figures in the left column). A slight increase of the electrons' number per UC in its turn makes these regions further from the quantum well regions (the 3rd row of the figure) and the square symmetry of the lattice becomes more obvious.

In Fig.\ \ref{Polarization_QD} spin polarization in the absence of (left column), and in the presence of (right column) the EPC in a QD SL is plotted for different values of the number of electrons and a magnetic flux per UC.
For large enough values of $N_{e}$ and $pq$ (see the 3rd row in the figure) spin polarization obviously reflects the square symmetry of the lattice. Note, that for all the considered values of parameters the spin-polarization, like the electronic density, is slightly smoothed out due to the EPC. Here, like will become evident later, we see
that the EPC is reducing the exchange part of the Coulomb interaction resulting in smaller spin gaps
and spin polarization.

The same as in Fig.\ \ref{Polarization_QD} but for the antidot structure and for other values of parameters $pq$ and $N_{s}$ is shown in Fig.\ \ref{Polarization_AQD}. Clear is a more pronounced effect of the lattice symmetry on the spin polarization as the 2DES now is never localized in the lattice cells.
\section{Magnetization}
%
\begin{figure}
      \includegraphics[width=0.23\textwidth,bb= 60 50 360 300]{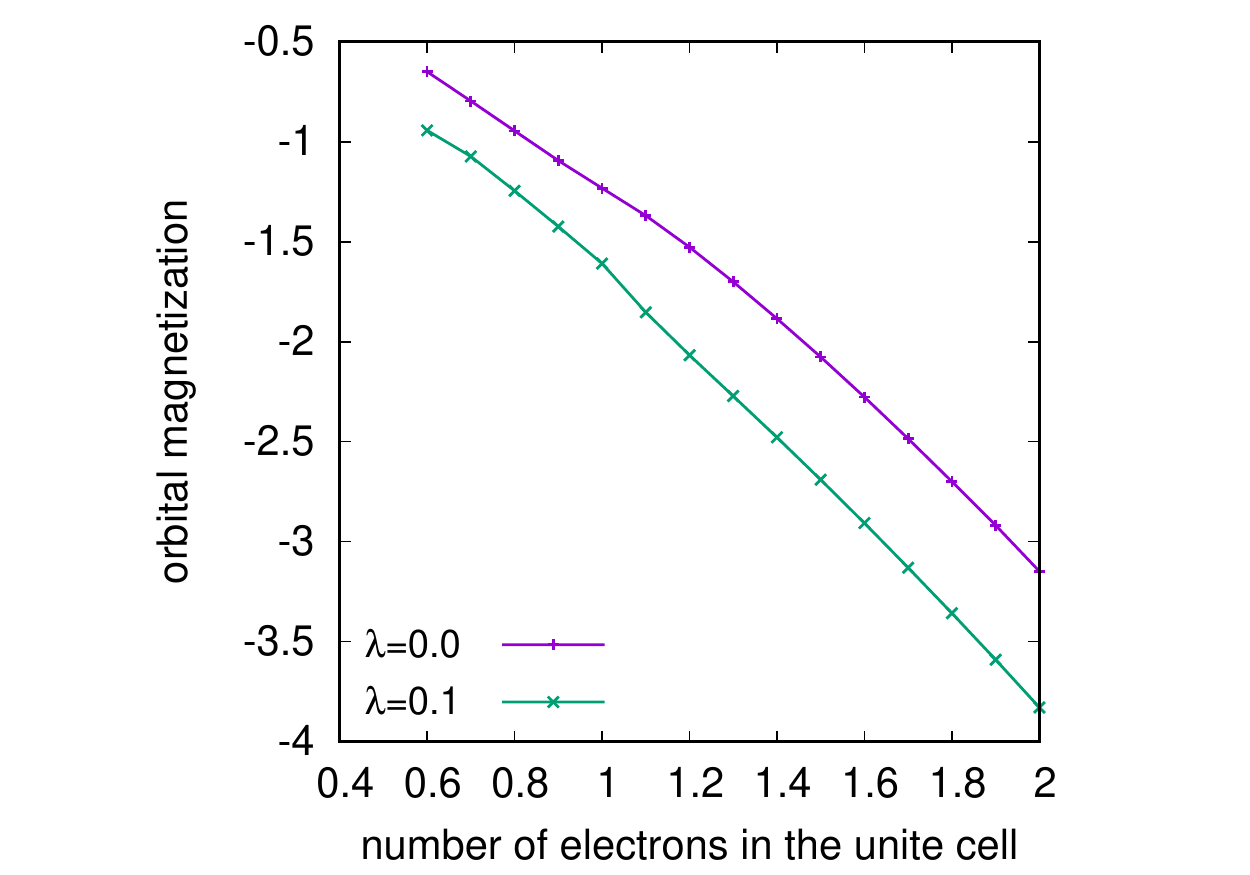}
      \includegraphics[width=0.23\textwidth,bb= 55 50 355 300]{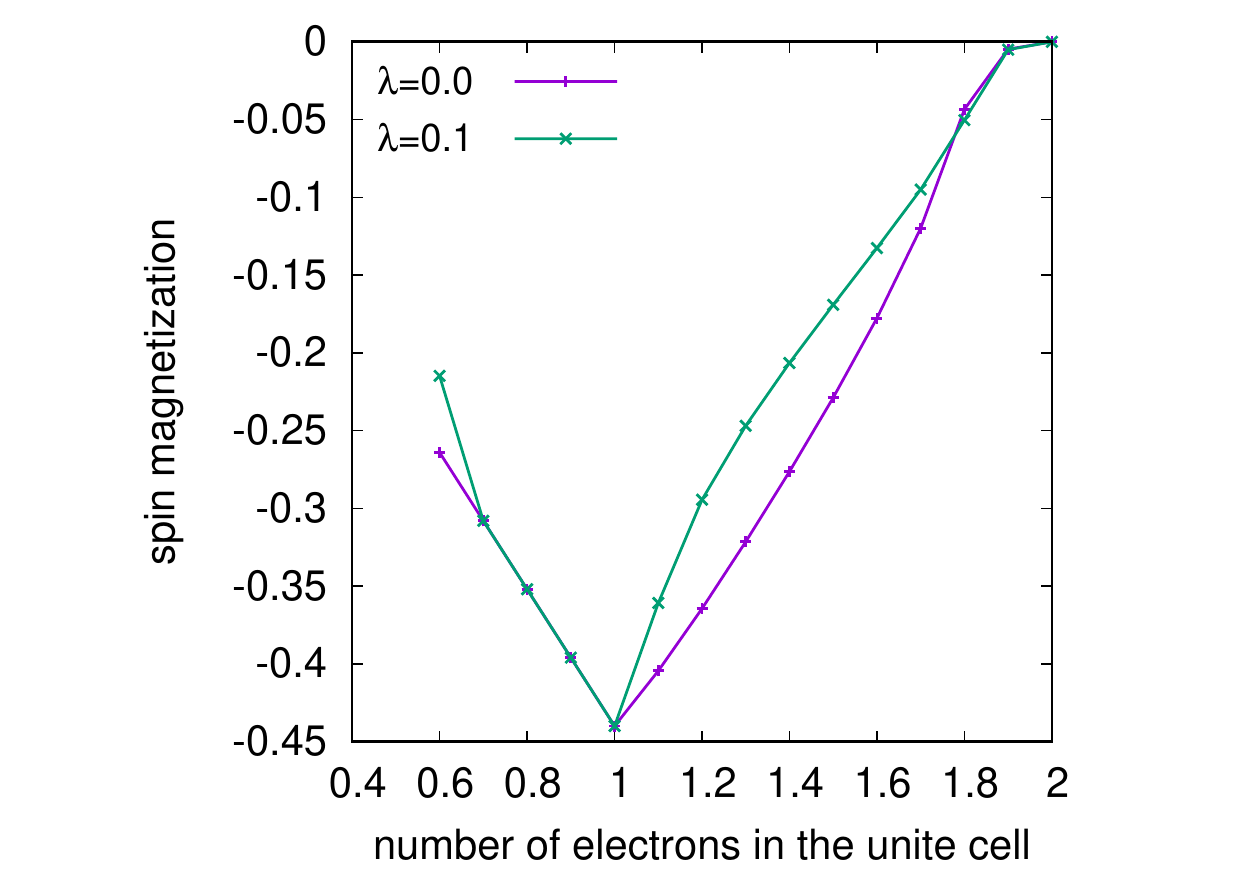}
      \label{MoD_La0La01_pq1_hwa1}
      \includegraphics[width=0.23\textwidth,bb= 55 50 355 300]{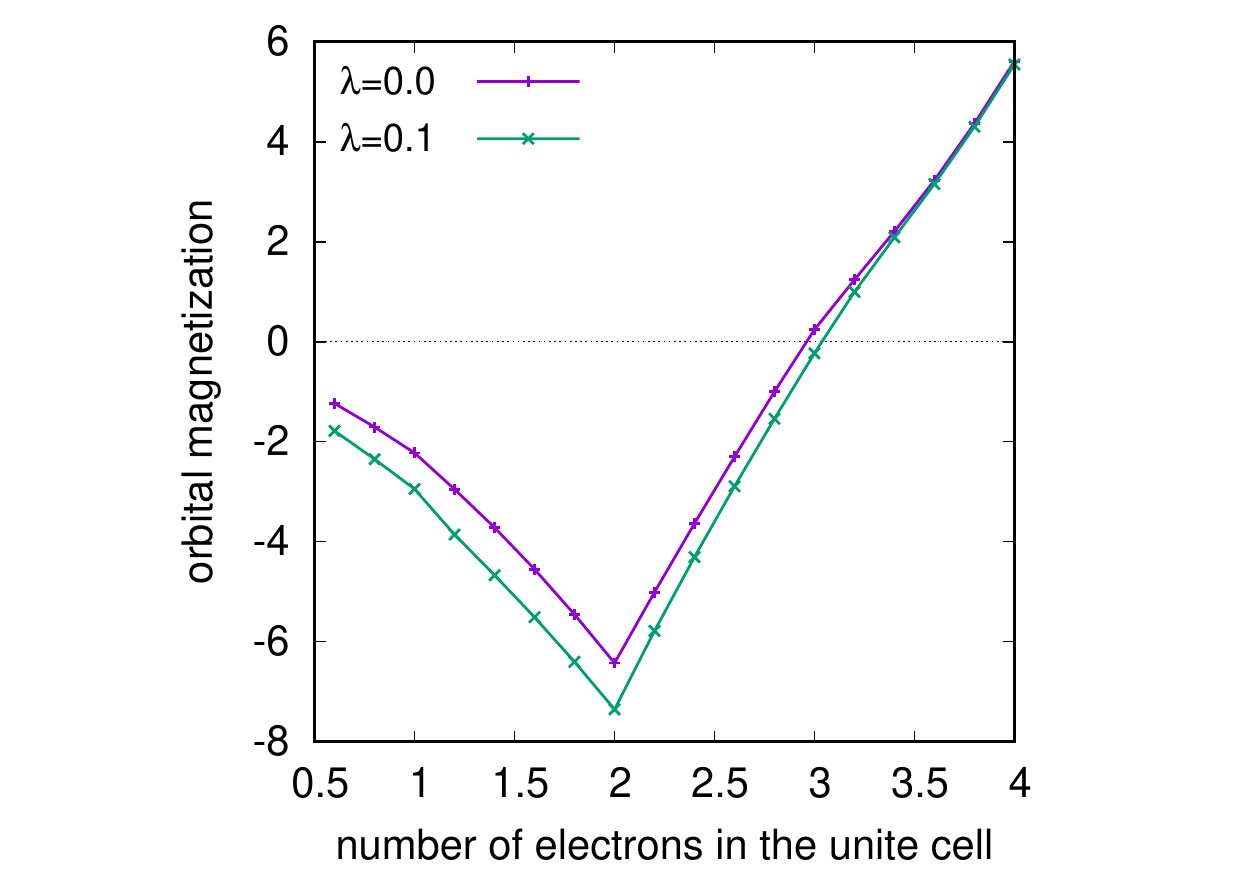}
      \includegraphics[width=0.23\textwidth,bb= 55 50 355 300]{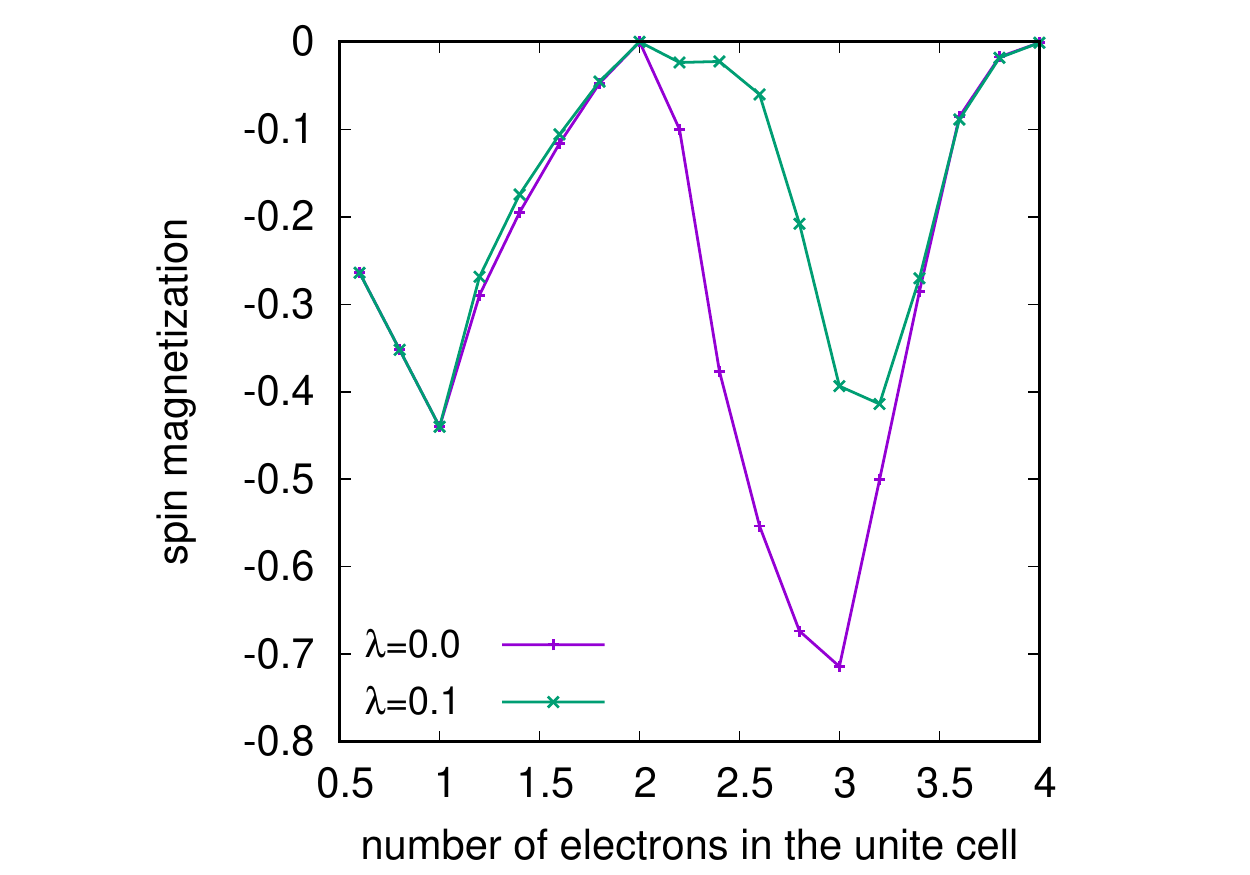}
      \label{MsD_La0La01_pq2_hwa1}
      \includegraphics[width=0.23\textwidth,bb= 55 50 355 300]{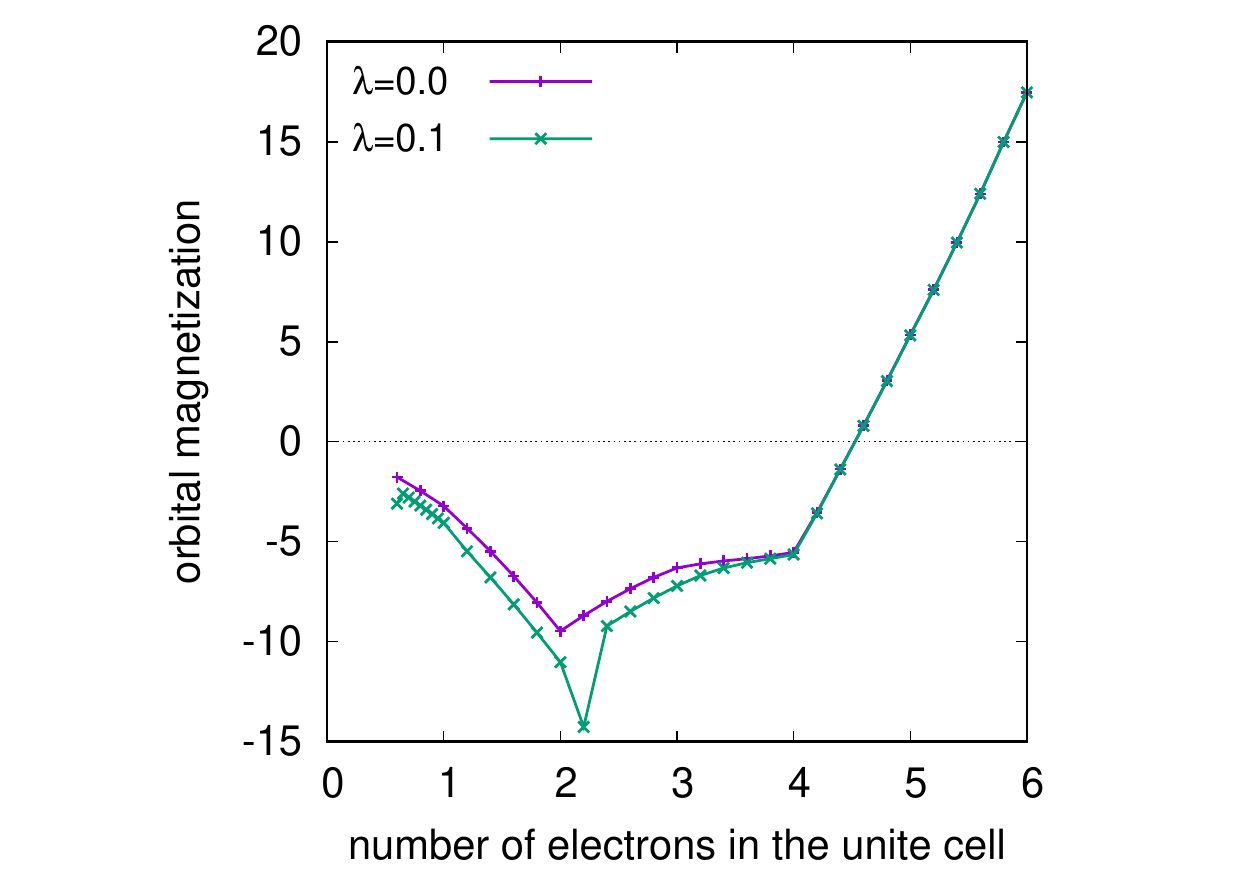}
      \includegraphics[width=0.23\textwidth,bb= 55 50 355 300]{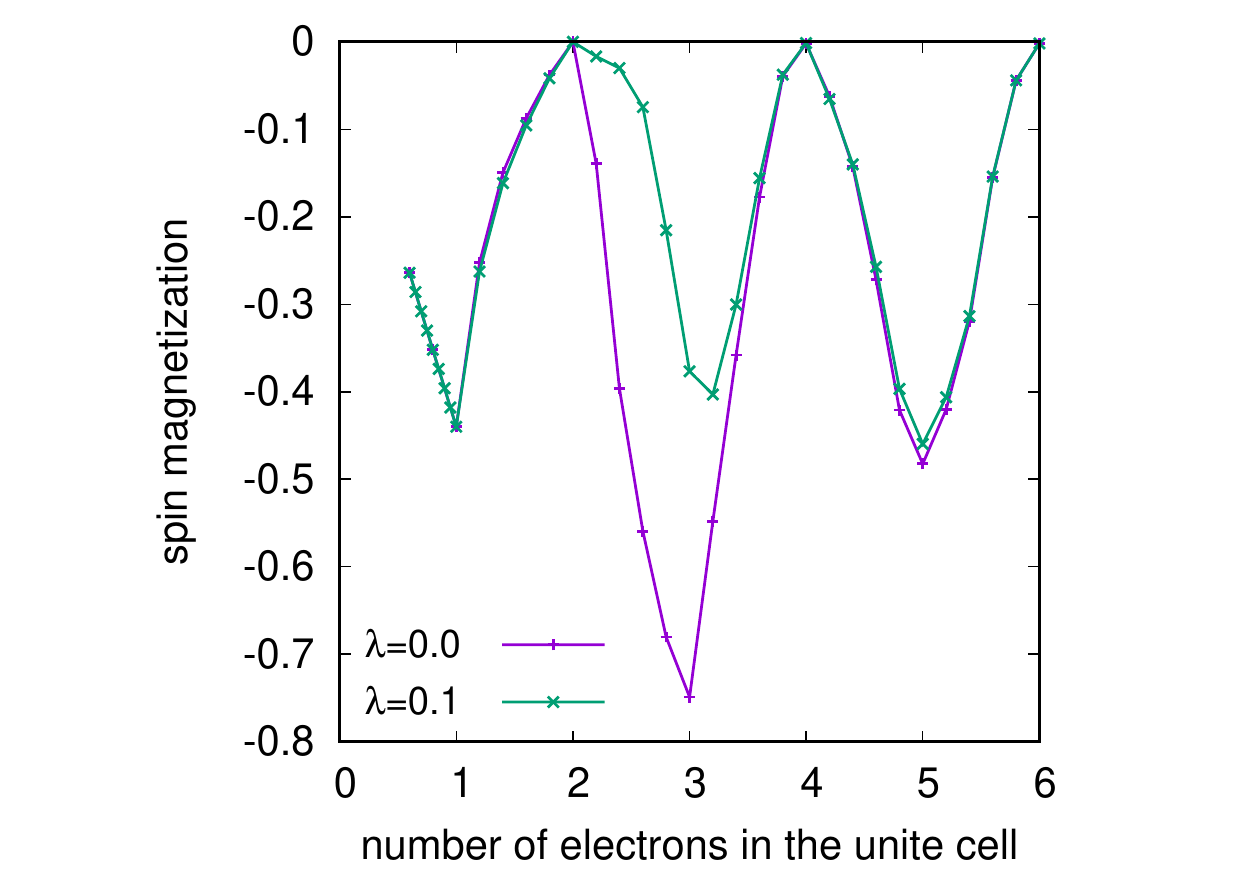}
      \label{MsD_La0La01_pq3_hwa1}
      \includegraphics[width=0.23\textwidth,bb= 55 20 355 300]{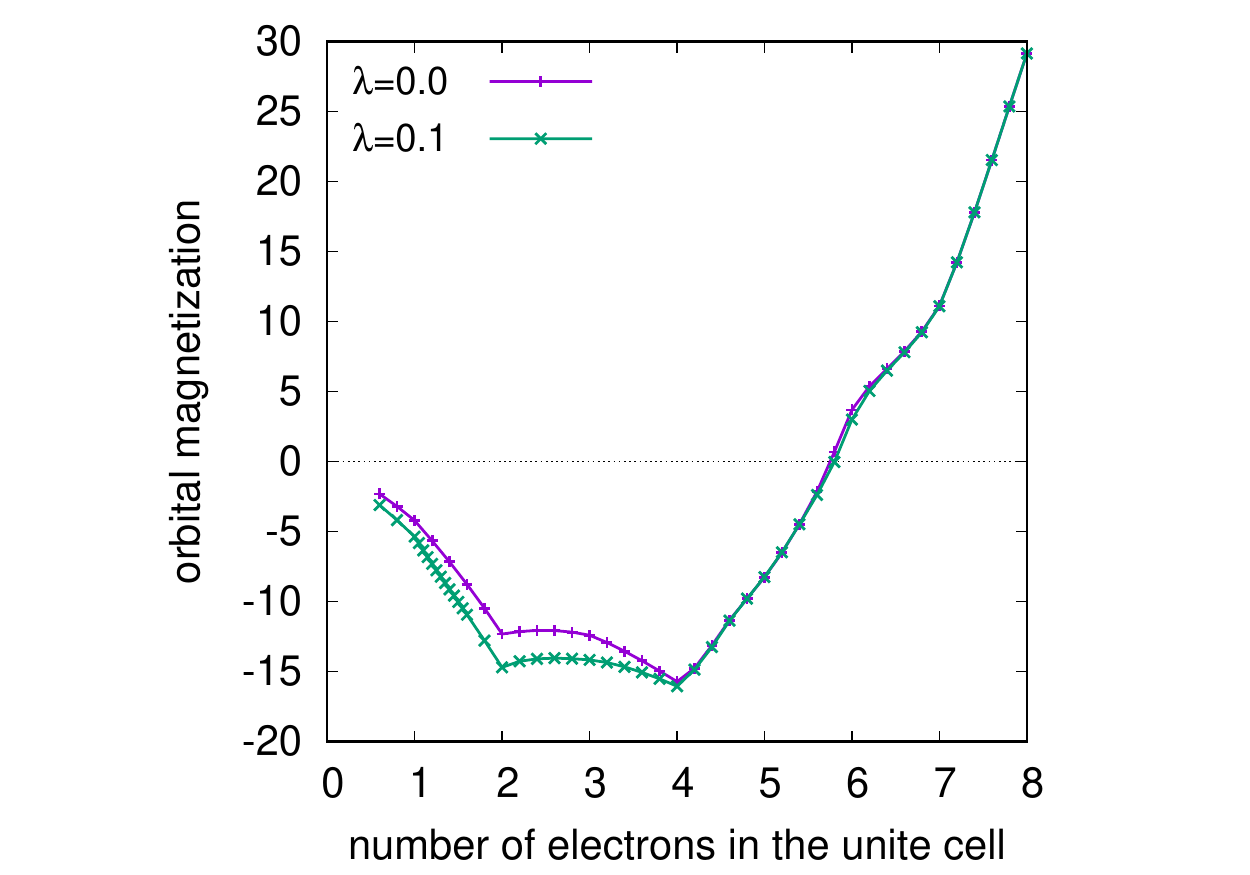}
      \includegraphics[width=0.23\textwidth,bb= 55 20 355 300]{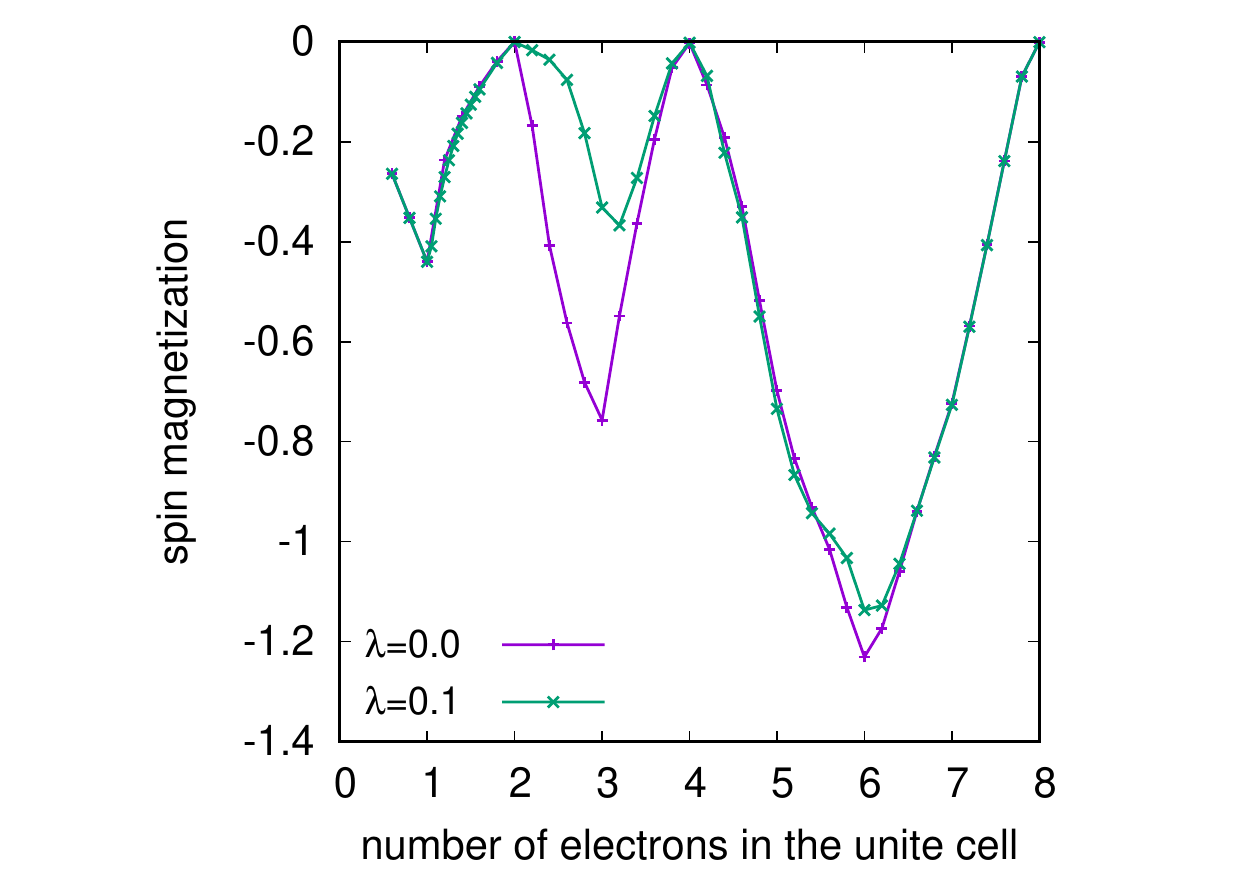}
      \label{MsD_La0La01_pq4_hwa1}
      \caption{Orbital (left column) and spin (right column) magnetization of a QD SL
      	in terms of $M_0 = \mu_\mathrm{B}^*L^2$.
      	The rows from the top to the bottom correspond to the values $pq=1$, $pq=2$, $pq=3$ and $pq=4$, respectively.}
      \label{MD}
\end{figure}

\begin{figure}
      \includegraphics[width=0.23\textwidth,bb= 55 50 355 300]{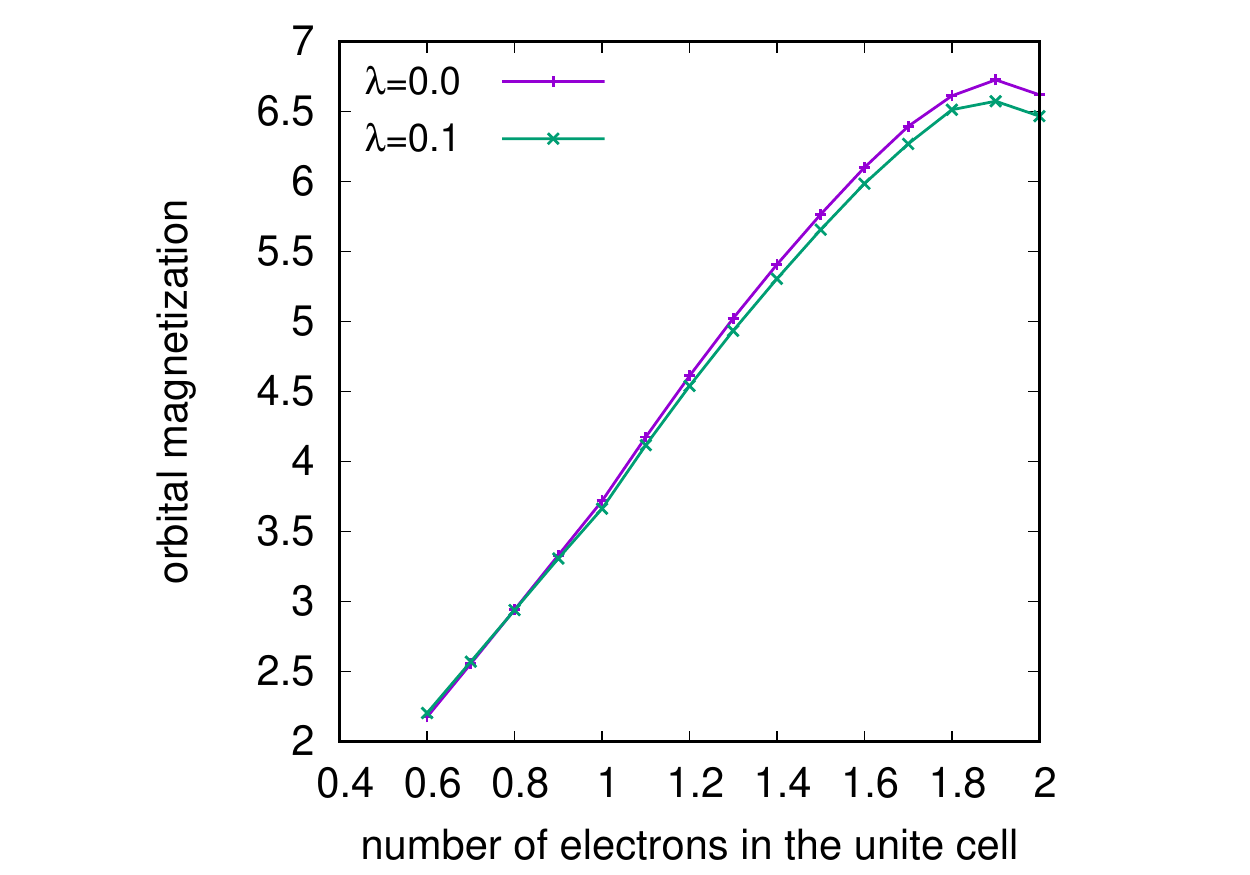}
      \includegraphics[width=0.23\textwidth,bb= 55 50 355 300]{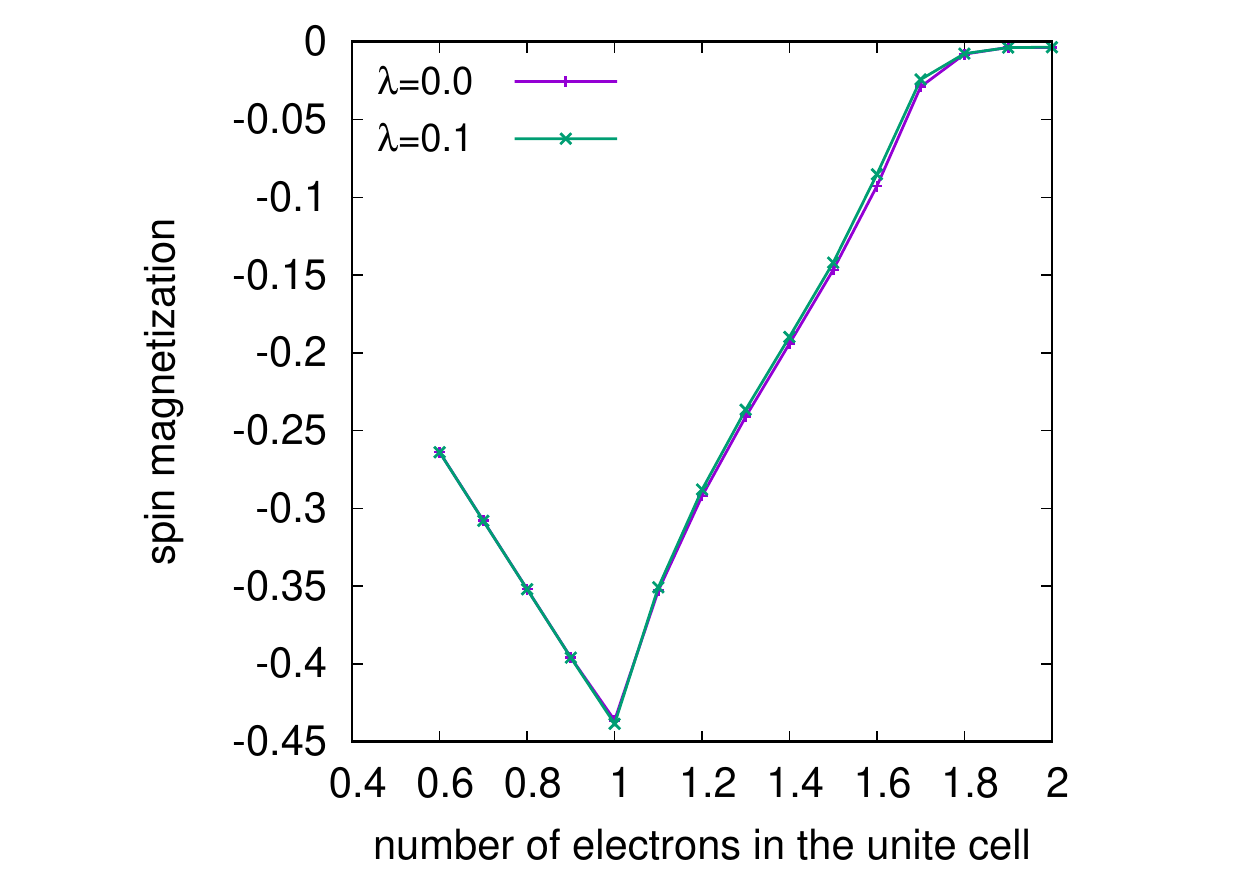}
      \includegraphics[width=0.23\textwidth,bb= 55 50 355 300]{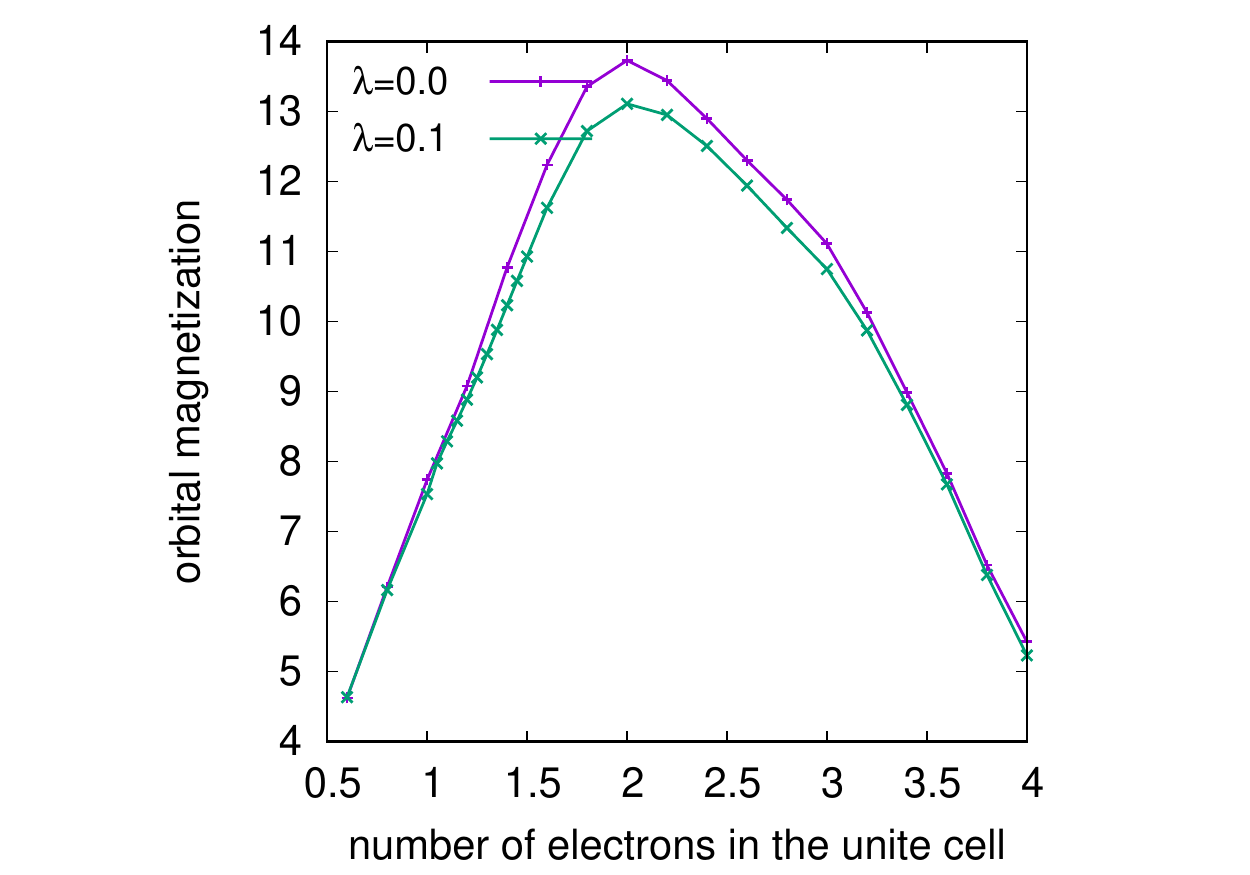}
      \includegraphics[width=0.23\textwidth,bb= 55 50 355 300]{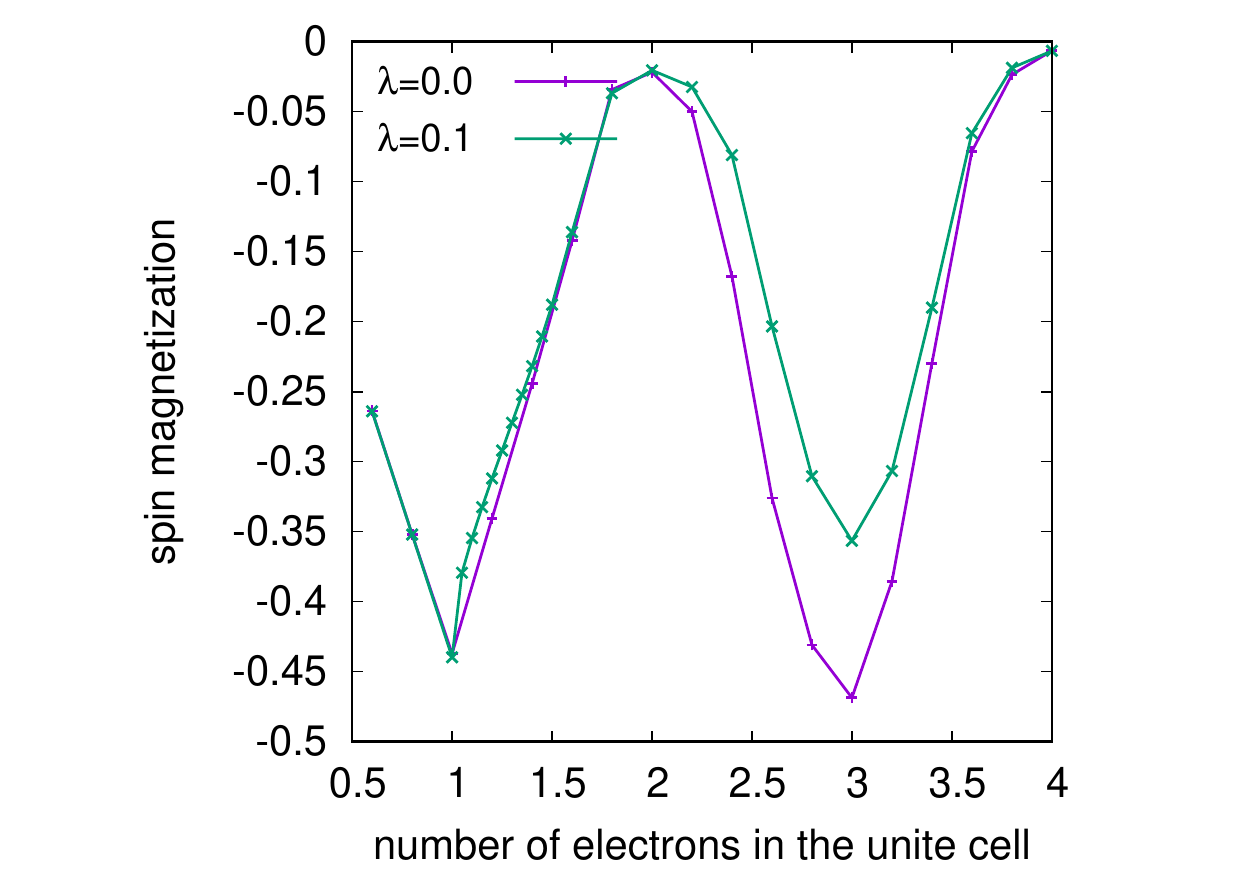}
      \includegraphics[width=0.23\textwidth,bb= 55 50 355 300]{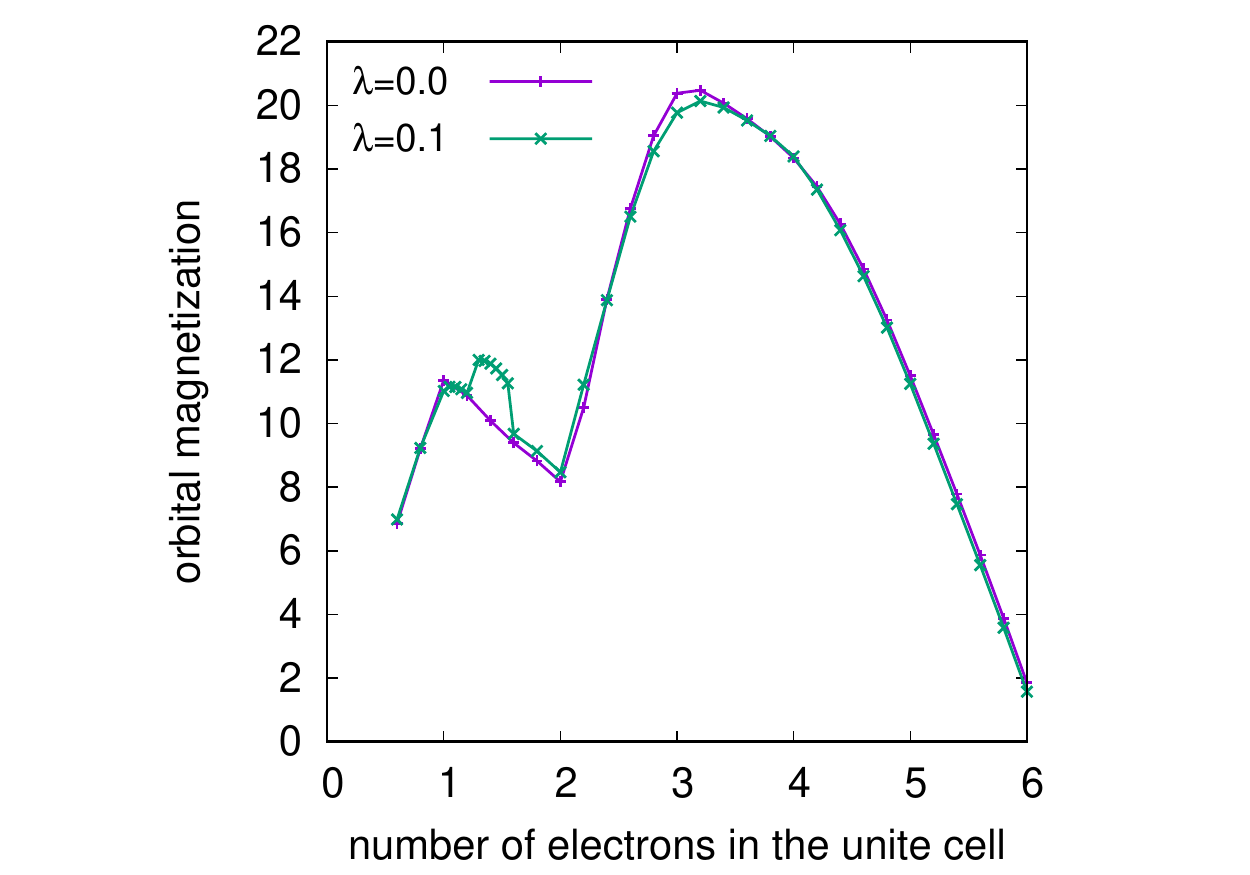}
      \includegraphics[width=0.23\textwidth,bb= 55 50 355 300]{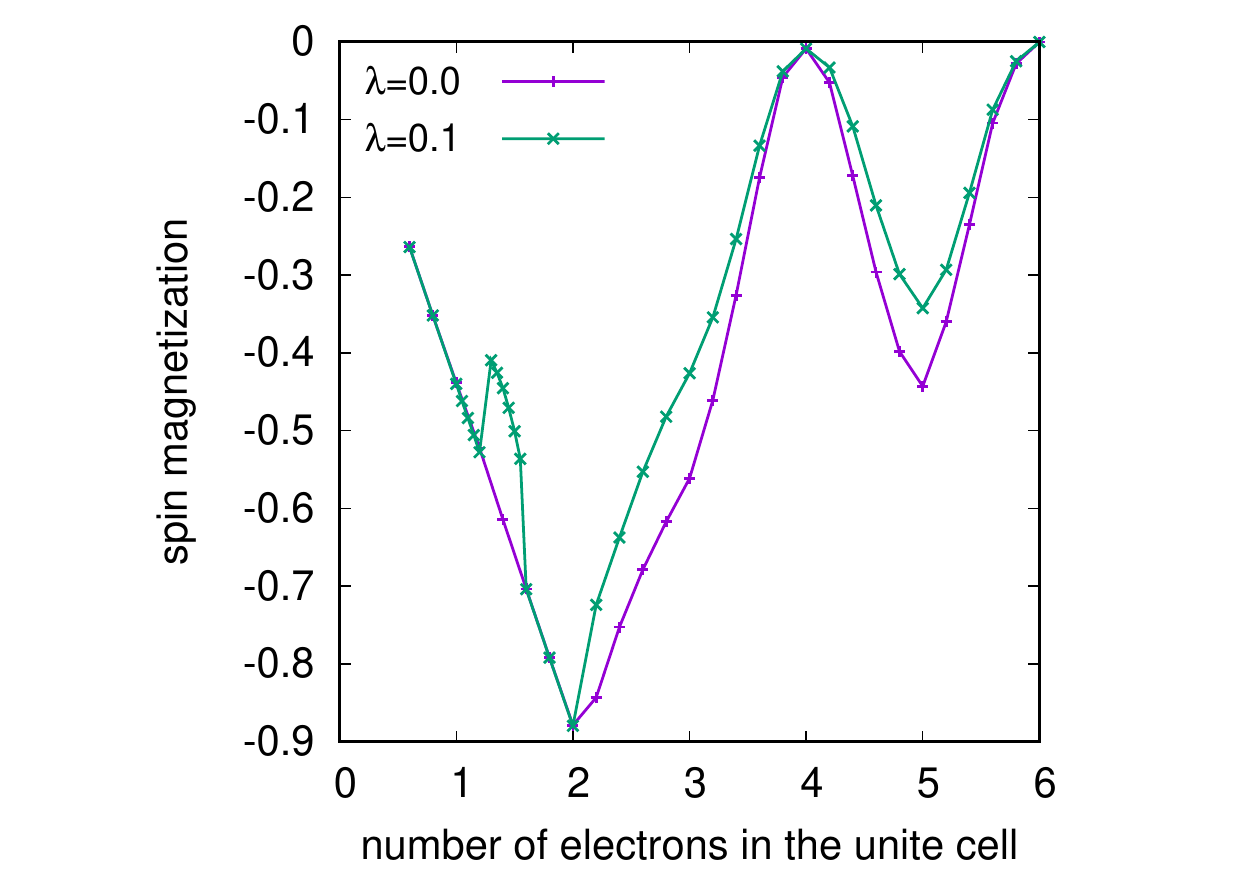}
      \includegraphics[width=0.23\textwidth,bb= 55 20 355 300]{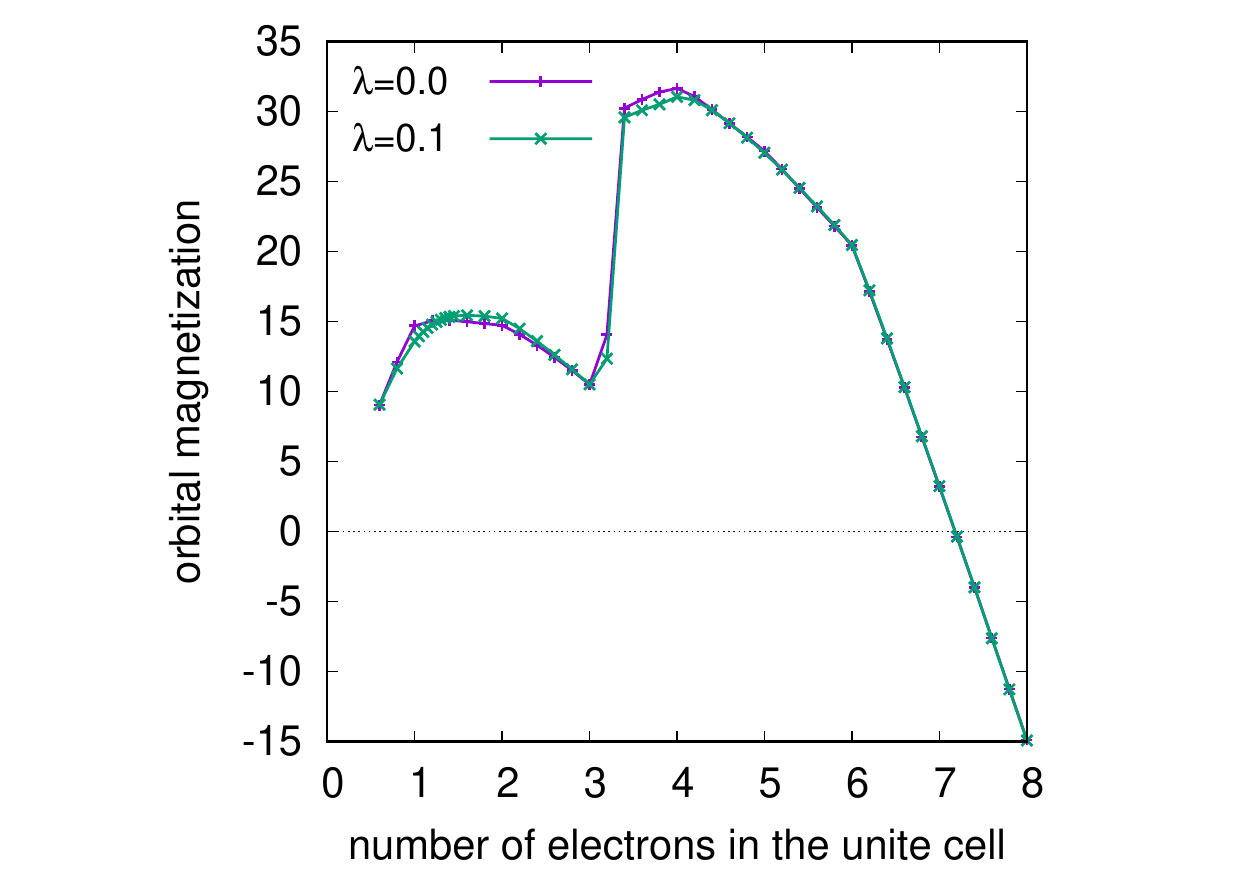}
      \includegraphics[width=0.23\textwidth,bb= 55 20 355 300]{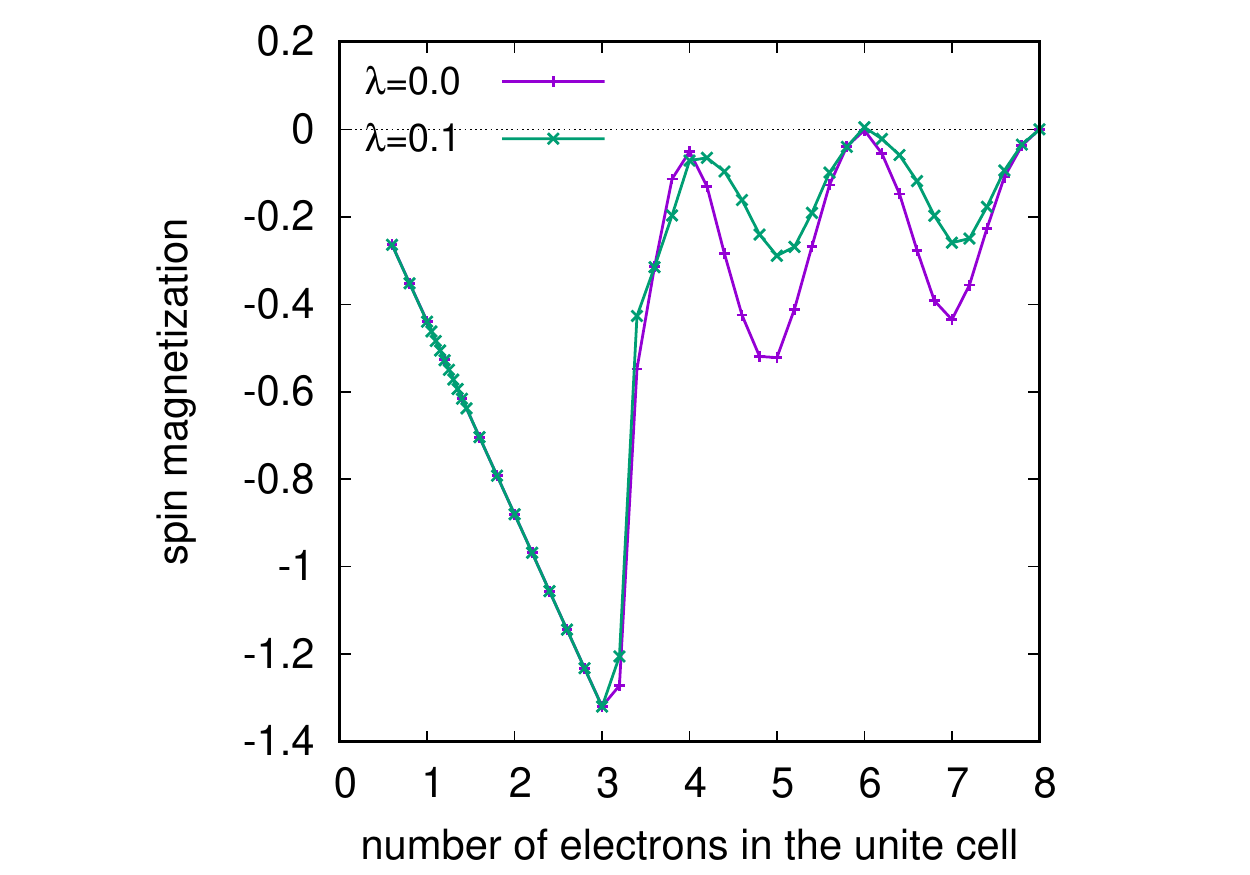}
      \caption{Orbital (left column) and spin (right column) magnetization of an AQD SL
      	in terms of $M_0 = \mu_\mathrm{B}^*L^2$.
      	The rows from the top to the bottom correspond to the values $pq=1$, $pq=2$, $pq=3$ and $pq=4$, respectively.}
      \label{MA}
\end{figure}
The orbital and the spin components of the magnetization of the system are accessible through the expression
\begin{equation}
\label{magnetization}
\begin{aligned}
M_{o}+M_{s}= & \frac{1}{2 c \mathcal{A}} \int_{\mathcal{A}} d \boldsymbol{r}\: (\mathbf{r} \times \mathbf{j}(\mathbf{r})) \cdot \hat{\mathbf{e}}_{z} \\
& -\frac{g^{*} \mu_{B}^{*}}{\mathcal{A}} \int_{\mathcal{A}} d \boldsymbol{r}\: \sigma_{z}(\mathbf{r}),
\end{aligned}
\end{equation}
where $\mathcal{A}=L^2$ is the area of a SL unit cell, $\mathbf{j}(\mathbf{r})$ is the mean
current density and $\sigma_{z}(\mathbf{r})$ is the mean spin density \cite{Gudmundsson1}.

In Fig.\ \ref{MD} the orbital (left column) and the spin (right column) magnetizations (in units of $\mu^{*}_\mathrm{B}/L^{2}$) for the QD SL are presented versus the number of electrons $N_{e}$ per UC. The purple curves correspond to no interaction with a photon cavity, while the green ones are for the presence of a EPC. In each case the maximum value of $N_{e}$ corresponds to the value of the average filling factor $\nu =2$.
As is seen from the figure the EPC leads to a stronger orbital magnetization (OM) for almost all the considered values of magnetic flux through the UC. This is a result of the smearing of electron charge, reflecting its polarizability,      over the UC, and hence, the weakening of the confinement
modifying the paths of the mean persistent charge current in the system. For relatively small values of the number of electrons per UC the system is diamagnetic (in the case of $pq=1$ it is the case for all the considered values of $N_e$), while for larger values of $N_e$, one can observe a paramagnetic behavior with positive magnetization. For $pq=2$ the turn around for the OM, when $\lambda=0$, starts from $N_e=2$, because a subband with angular momentum opposite to the magnetic field starts to be filled (the 2nd and the 3rd figures in the left column). It is noteworthy that the diamagnetic effect is more prominent for $pq=3$ and $pq=4$ (the 3rd and the 4th figures in the left column) because of higher values of angular momenta of the populated subbands. Note, that for large enough values of $N_e$ the EPC has almost no effect on the OM. This is because of a significant contribution to the OM of the electronic subbands with higher energies which are weakly localized in QDs and hence their spatial distributions is not strongly affected by the cavity field.
Interestingly, when $pq=3$ one can observe for cavity-coupled electrons a slight deviation from $N_e=2$ of the onset of a paramagnetic contribution to the OM.

The spin magnetization (SM) undergoes oscillations versus the electrons number per UC and becomes zero at even integers of $N_e$ for $pq=1$, $2$ and $3$ (the first three figures in the right panel). This situation corresponds to the grouping of electrons in couples with mutually opposite spin states along with the filling of energy subbands. In contrast, the local maxima for the magnitude of SM for above mentioned values of $pq$ correspond to odd number of electrons in each UC, which corresponds to totally not compensated spin of one electron.
For stronger magnetic fields the strong exchange interaction leads to a subband reorganization.
The signature of this is seen for $pq=4$ (the 4-th figure in the right panel) where SM has its minimum at $N_e=6$ and becomes zero at $N_e=8$.
As is seen from the 2-nd, 3-rd and the 4-th figures in the right column, the SM weakens significantly due to the EPC around $N_e=3$. To understand this effect we should keep in mind that the interaction with the cavity photons leads to the occupation of quantum states with opposite spins which in its turn results in a weaker exchange energy. This results in a weaker spin polarization and smaller magnitude for the SM.

In Fig.\ \ref{MA} the OM (left column) and the SM (right column) are presented (in units of $\mu^{*}_{B}/L^{2}$) for the AQD SL versus $N_{e}$. As is obvious from the figure, for $pq=1$, $pq=2$ and $pq=3$ the MO is positive over all the range of $N_{e}$. In this regard an AQD SL behaves rather like a structure composed of quantum rings, where the role of the states with zero angular momentum is suppressed due to the repulsion of electrons from the UC center (for high enough values of magnetic field the contribution of negative angular momentum in the ground state energy becomes prominent). With increasing $N_{e}$ the population of the states with negative momentum increases leading to an increase of the OM. The further increase of $N_{e}$ leads to filling of higher states with positive angular momentum. These two processes are being periodically exchanged leading to an oscillatory behavior of the OM with increasing $N_{e}$ (the left column in Fig.\ \ref{MA}). This behaviour of the OM is quite similar to, but not exactly the same as for a quantum ring in a transverse magnetic field.

The spin magnetization of the two systems can be used to obtain further knowledge about them.
For the dot array there is always a local minimum at $N_e=1$. For the AD SL this is only true at low magnetic
flux. Not for $pq=3,4$. The reason is that the Coulomb exchange is generally stronger in the AD SL, where the
electron density is continuous, but not localized in dots. The stronger Coulomb exchange
leads to larger spin splitting
and thus a rearrangement of the lowest energy spin bands. This effect is also seen in the OM, remember,
the hill in the AD SL is in the center of each UC. For low $N_e$ this hill is weakened by the EPC, but the
increased direct Coulomb interaction counteracts this effect and the generally strong Coulomb exchange
force too, by pulling the electrons away from the hill region. This explains the weaker effects of the EPC
on the MO for the AD SL. This picture is strengthened by the fact that the charge modulation by the
AD potentials is less, than by the QD potentials.

In general, one can observe a slight decrease in the magnitude of the OM due to the EPC, which is the result of the smearing effect of the EPC (like in the case of a QD SL), or the polarization, leading to a stronger penetration of electrons into the UC center (in contrast to the case of a QD SL). However, an opposite behavior of the OM is observed for $pq=3$ and $N_e \simeq 1 - 1.6$. This result is a consequence of filling of states with higher energy and negative angular momentum resulting in a paramagnetic contribution to the total OM. It is noteworthy, that the above mentioned increase in the magnitude of the OM is accompanied with the decrease in the magnitude of the SM for the same values of $pq$ and $N_e$, which is in accordance with the fact that the system tends to minimize its total energy.

In light of the description of the role of the Coulomb exchange for the AD SL one paragraph
above we can understand this change in the MO and the SM for $pq=3$ around $N_e \simeq 1 - 1.6$ as the
``come back'' of the minimum for the SM at $N_e=1$ promoted by the EPC. The reason being the delicate
balancing of the exchange and correlation terms of the Coulomb and the electron-photon interactions, but
in light of our computational approach, we have to admit that it is difficult to maintain that the
present DFT and QEDFT approach can totally accurately describe this delicate balance.

\section{CONCLUSIONS}
Quantum electrodynamical density functional
theory is applied to an electron gas in a superlattice of square symmetry placed in a far-infrared photon cavity and subjected to an external transverse and homogeneous magnetic field. The cavity field assumed to be created by two parallel plates providing isotropic photon polarization in the superlattice plane. Comparative study for quantum dot- and antidot lattices is presented. The Coulomb interaction between the electrons and the electron-photon coupling are treated self-consistently. We explore how the interplay between the Coulomb interaction and the electron-photon coupling affects the electron density, the magnetization and the spin polarization of the system.
We find that both the orbital and the spin magnetization of the electron system change in nontrivial ways, depending on the number of electrons and the magnetic flux in each unit cell of the superlattice. For a low number of electrons in a unit cell, we observe a diamagnetic behavior for the orbital magnetization in the quantum dot structure and a paramagnetic behavior in the antidot structure. However, with an increasing number of electrons the orbital magnetization undergoes nontrivial changes connected with the role of the exchange and correlation forces on the occupation of higher energy levels. For a quantum dot structure the cavity field strengthens the diamagnetic behavior of the electron gas, while for an antidot structure the electron-photon coupling results in weakening of the paramagnetic effect. The spin magnetization reveals oscillations with increasing number of electrons per a unit cell, both for the quantum dot and the antidot structures. For comparatively small values of the magnetic field flux per unit cell the spin magnetization is zero for even integer values of electrons per unit cell, which indicates s pairing of electrons with opposite spins in each energy subband. For larger values of magnetic flux the magnitude of spin magnetization may differ from zero and pass trough a maximum value when the number of electrons is an even integer. This result reflects an interplay between the magnetic field and the spin polarization caused by the exchange and correlation energy of the electrons. Generally, the electron-photon coupling leads to a weakening of the spin magnetization.
As the number of electrons per unit cell of the lattice increases the electron-photon interaction reduces the exchange forces that would otherwise promote strong spin splitting for both the dot
and the antidot array.

\vspace{0.5cm}

\section{ACKNOWLEDGMENTS}
This work was financially supported by the Research Fund of the University of Iceland, and the Icelandic Infrastructure Fund. The computations were performed on resources provided by the Icelandic High Performance Computing Center at the University of Iceland. V. Mughnetsyan and V.Gudmundsson acknowledge support by the Armenian State Committee of Science (grant No 21SCG-1C012). V. Mughnetsyan acknowledges support by the Armenian State Committee of Science (grant No 21T-1C247). V. Moldoveanu acknowledges financial support from the Core Program of the National
Institute of Materials Physics, granted by the Romanian Ministry of Research, Innovation and Digitalization under the Project PC2-PN23080202.
%

\end{document}